\documentclass[namedreferences]{kluwer}
\usepackage{graphicx}

\begin{document}
\begin{article}

\begin{opening}

\title{Bulk Flow Regimes and Fractional Flow in 2D Porous Media by Numerical Simulations}

\author{HENNING ARENDT \surname{KNUDSEN$^{1,2}$}\email{arendt@phys.ntnu.no}}
\author{EYVIND \surname{AKER$^{3,4,5}$}\email{eyvind.aker@westerngeco.com}}
\author{ALEX \surname{HANSEN$^{1,2}$}\email{Alex.Hansen@phys.ntnu.no}}

\institute{$^1$Department of Physics, Norwegian University of Science and 
Technology, NTNU, NO-7491 Trondheim, Norway}
\institute{$^2$Nordita, Blegdamsvej 17, DK-2100 Copenhagen, Denmark}
\institute{$^3$Department of Physics, University of Oslo, NO-0316 Oslo, Norway}
\institute{$^4$Niels Bohr Institute, Blegdamsvej 17, DK-2100 Copenhagen, Denmark}
\institute{$^5$WesternGeco, Schlumberger House, Solbr{\aa}vn 23, NO-1372 Asker, Norway}

\runningtitle{2D BULK FLOW SIMULATIONS}
\runningauthor{HENNING ARENDT KNUDSEN ET AL.}

\begin{abstract} 
We investigate a two-dimensional network simulator that model the dynamics of two-phase immiscible bulk flow where film flow can be neglected. We present a method for simulating the detailed dynamical process where the two phases are allowed to break up into bubbles, and bubbles are allowed to merge together. The notions of drainage and imbibition are not adequate to describe this process since there is no clear front between the fluids. In fact, the simulator is constructed so that one can study the behaviour of the system far from inlets and outlets, where the two fluids have been mixed together so much that all initial fronts have broken up. The simulator gives the fractional flow as a function of the saturation of each of the fluids. For the case of two fluids with equal viscosity, we classify flow regimes that are parametrized by the capillary number.
\end{abstract} 

\end{opening}

\section{Introduction}
Two-phase displacements have been studied by experimental work \cite{CW85,MFJ85,LTZ88}, numerical simulations \cite{KL85,LTZ88,BK90}, statistical models \cite{WS81,WW83,P84} and differential equations \cite{BC98} during the past decades. Real-world motivations for work in this field are oil-recovery and hydrology. The complex nature of the problem makes it worthwhile to approach the problem in many ways.

In this paper we present a simulator that provides fractional flow curves as a function of saturation in a steady-state situation without boundaries. Much work in the field of fractional flow has been done by means of solving differential equations, see for example \inlinecite{BC98} and references therein. There has also been numerical work, see \inlinecite{R90}. The classical work in this field is that of \inlinecite{BL42}. See also \inlinecite{D92} for a broader presentation.

Since we are dealing with a closed system without boundaries and no clear front, the notions of drainage and imbibition are not adequate to describe the process in this model system. We have therefore made a new classification of the flow regimes; viscous force-dominated regime, viscous and capillary interaction regime, and capillary force-dominated regime. We refer to them as bulk flow regimes, and they are characterized by different behaviour of the fractional flow curves as a function of saturation. The two first are also found to be history independent regimes, whereas the third regime shows history dependence. This paper is primarily concerned with the history independent regimes.

This simulator is based on a simulator for drainage \cite{AMHB98,AMH98}. The drainage simulator was able to simulate successfully the temporal evolution of fluid pressure in three regimes of drainage invasion; viscous fingering, capillary fingering and stable displacement \cite{LTZ88}. The presented model is quite detailed and requires substantial computational resources for system sizes above $100\times 200$ lattice points. Therefore, very large systems are inconvenient to work with, at least on single CPU computers. It is interesting to extend the model to three dimensions. The problem is suitable for parallel computers, which provide a means to study those systems.

The main change in the model presented here is the boundary conditions. Instead of simulating an invasion process, we use biperiodic boundary conditions. The volume of the two fluids are conserved during a simulation. After a time the simulation reaches a steady-state from which intrinsic properties of the system can be found. This gives information on how the fluids are distributed within pore space and the corresponding flow properties, {\it e.g.}, the fractional flow. The closest experimental study is that of Avraam and Payatakes (\citeyear{AP95a,AP95b,AP99}). In etched glass networks they observe complex bubble dynamics far from inlets and outlets.

The paper is organized as follows. Sections \ref{sec:geometry} to \ref{sec:menisci} describe the model, {\it i.e.}, the geometrical aspects, the flow conditions, and the moving rules for menisci at nodes, respectively. This model is based on the model by Aker {\it et al.} which is reported on in great detail in \cite{AMHB98}. We will repeat only briefly the aspects of the model that are unchanged, and provide details where we have made changes. Finally, in section \nolinebreak \ref{sec:simulations} we provide simulation results, and section \nolinebreak \ref{sec:discussion} contains some concluding remarks.

\section{The Representation of the Porous Medium and the Geometry of the Fluid Interfaces}
\label{sec:geometry}
The network model presented here is based on the one presented thoroughly in Aker {\it et al.} (\citeyear{AMHB98,AMH98}). We will refer to that as the {\it mother model}. The porous medium geometry and the possible fluid interface configurations are essentially unchanged so only a short r\'esum\'e is given here.

\begin{figure}
\centering
\begin{tabular}{ll}
\includegraphics[height=4.5cm]{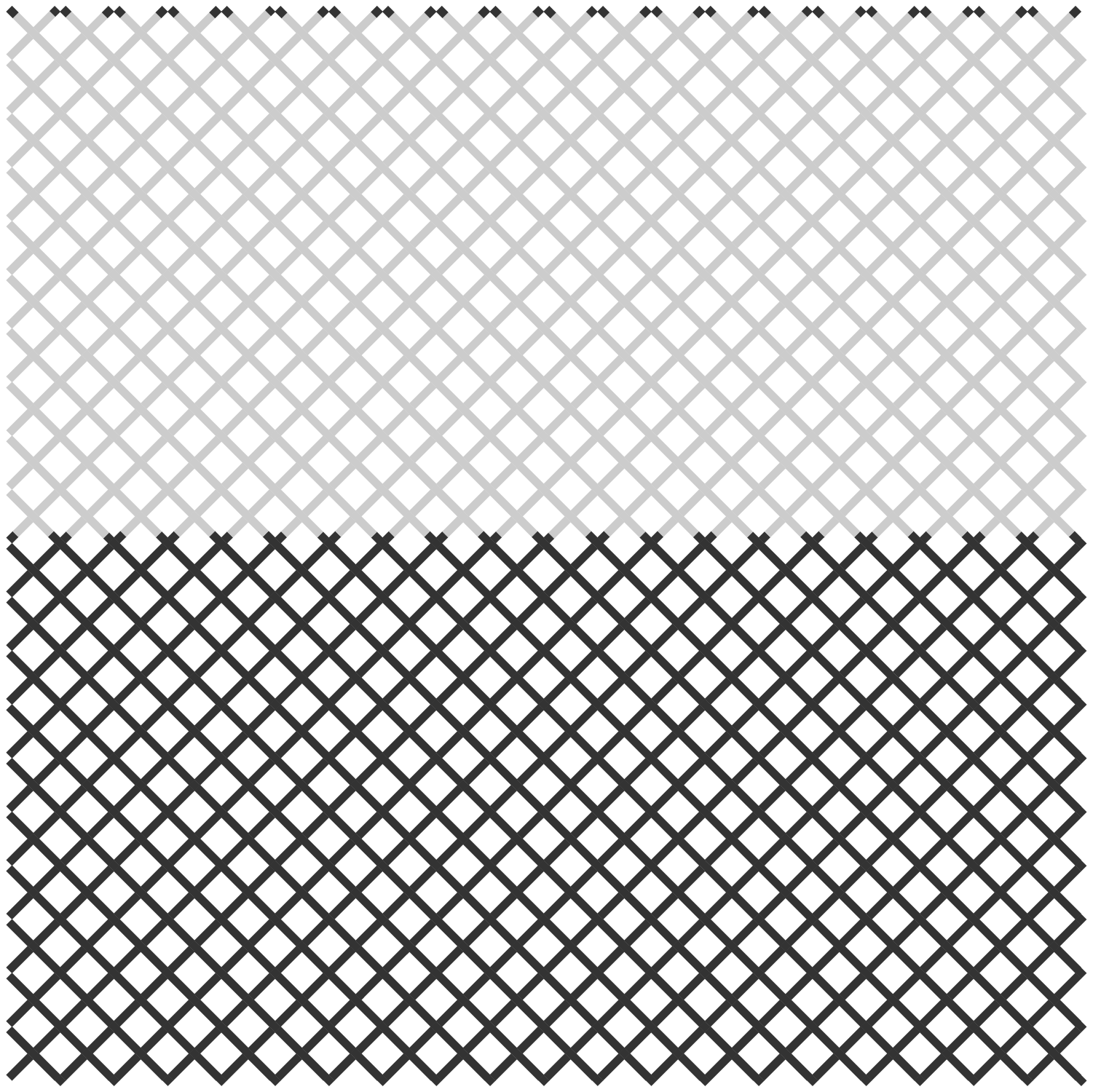} & 
\includegraphics[height=4.5cm]{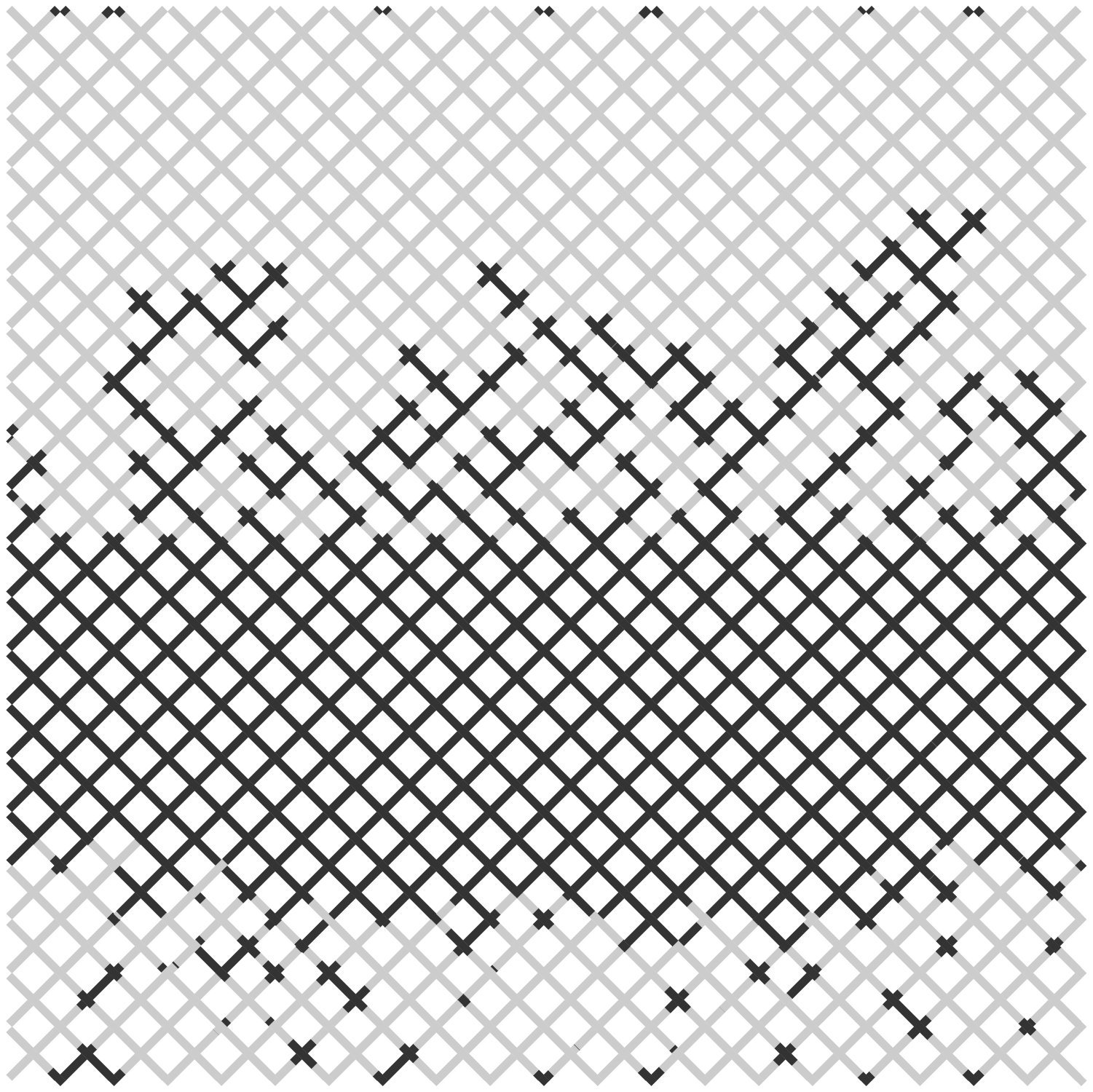} \\
(a) 0 s (0)  &
(b) 14.1 s (2000)  \\  &  \\
\includegraphics[height=4.5cm]{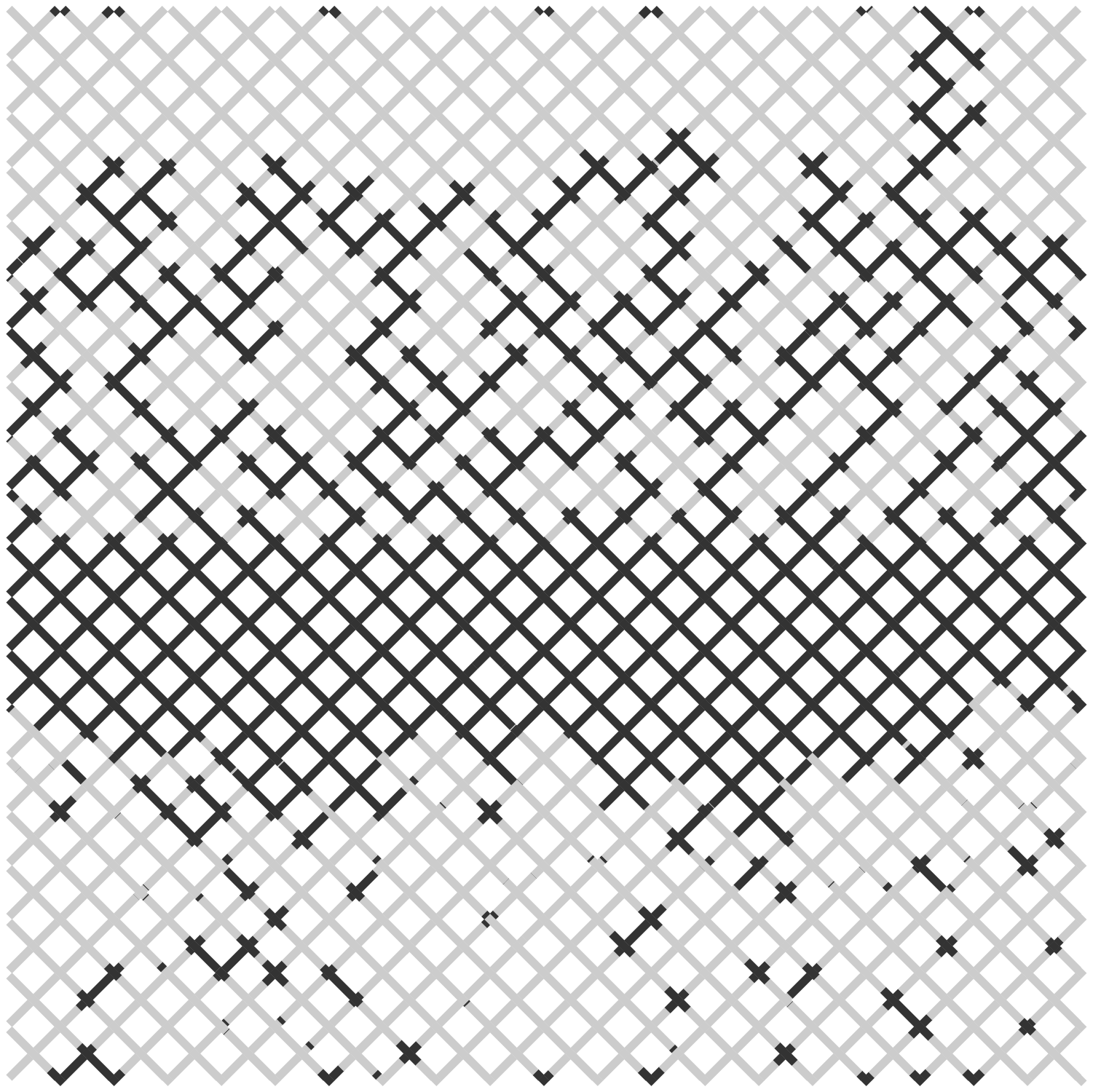} & 
\includegraphics[height=4.5cm]{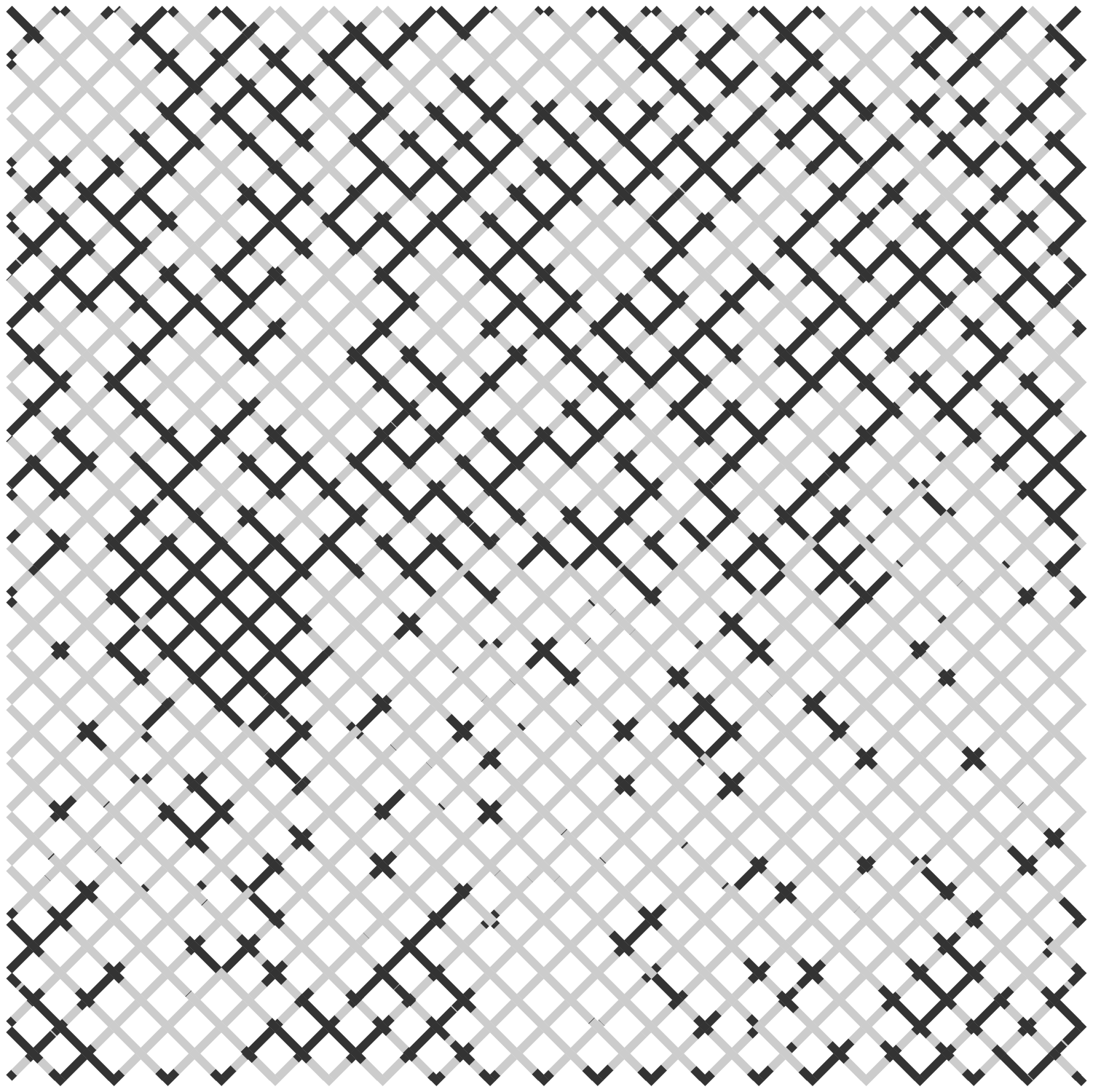} \\
(c) 24.0 s (4000) &
(d) 37.3 s (7000) \\  &  \\
\includegraphics[height=4.5cm]{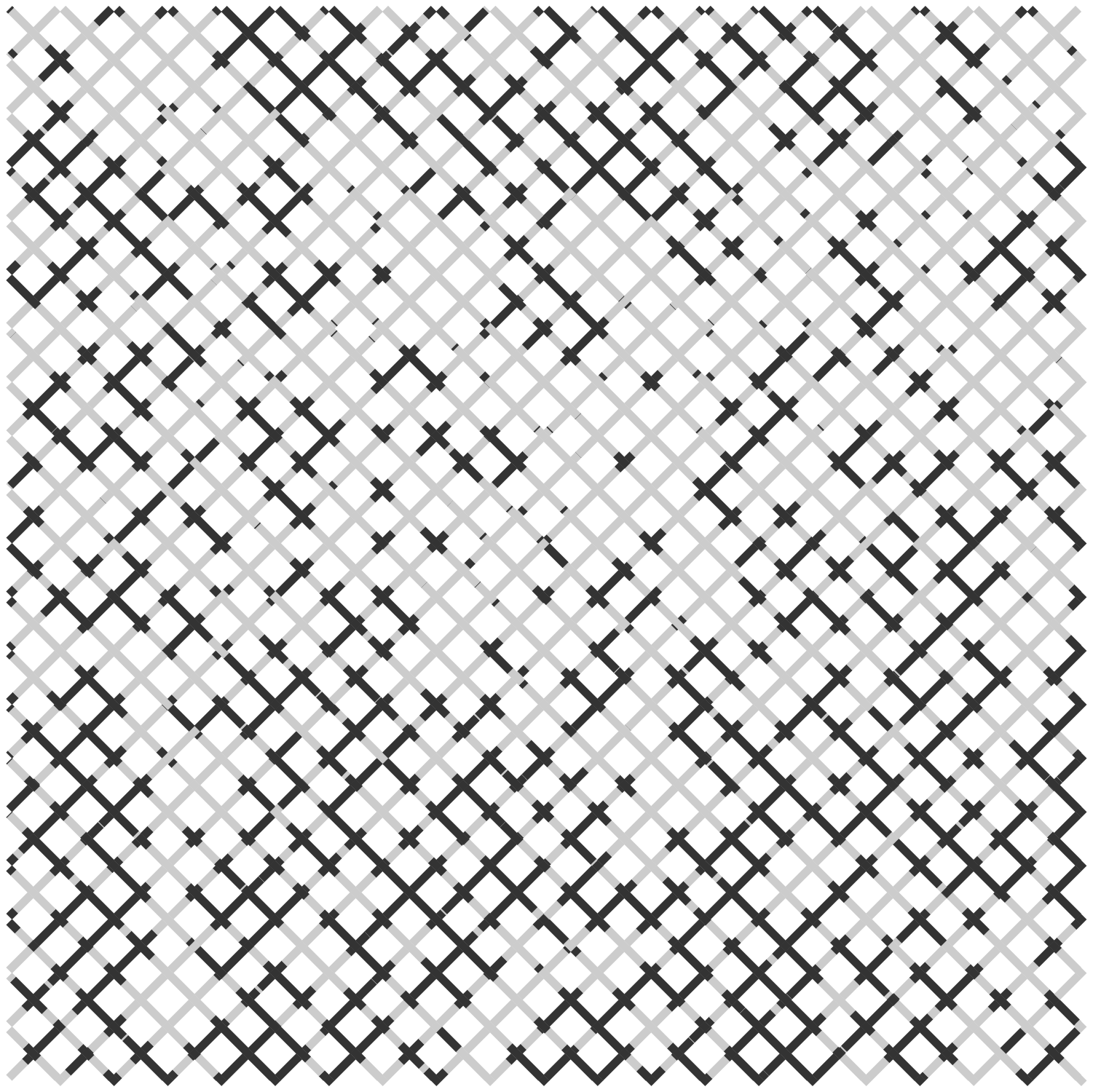} & 
\includegraphics[height=4.5cm]{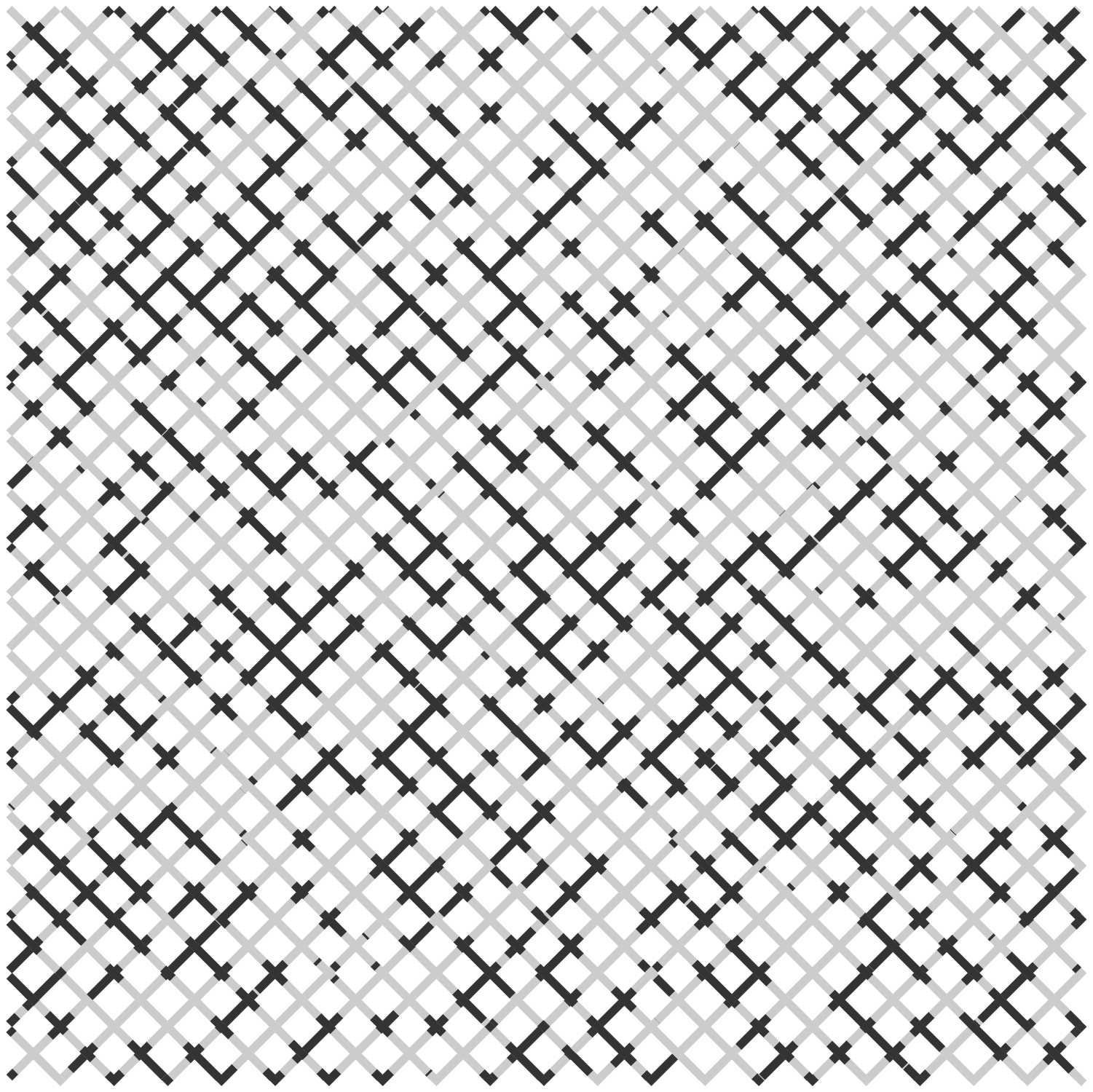} \\
(e) 59.3 s (12000) &
(f) 90.1 s (20000)
\end{tabular}
\caption{\textsl{Shows the time evolution of a $20\times 40$ nodes network. The nonwetting fluid is black, the wetting is gray. The illustrations are schematic in the sense that the variations in node positions and tube radii are not shown. The flow direction is from bottom to top, and the boundary conditions are biperiodic. The elapsed times are given in seconds and the numbers in the parentheses are the corresponding number of time steps. The nonwetting saturation is $S_{\rm nw}=50.2\%$, and the capillary number is $C_{\rm a}=1.0\times 10^{-3}$. (a) Initial fluid distribution. (b)-(e)  Transient period where pieces of the initial configuration can still be seen. (f) Steady state flow. Note, in particular, in (b) the instability of drainage in front of the nonwetting region and the stability of the imbibition process behind this region. In (f) the concepts of drainage and imbibition have lost their meaning.}}
\label{fig:networks}
\end{figure}

The porous medium is represented by a square lattice of cylindrical tubes, tilted $45^{\circ}$ with respect to the overall flow direction. Schematic network illustrations are found in Figure \ref{fig:networks}, where the flow direction is from bottom to top. The volume of throats and pores is contained in the tubes. The points where four tubes meet are referred to as nodes. Randomness is incorporated in the system by allowing the position of each node to be randomly chosen within the interval of plus minus thirty percent of the lattice constant away from its respective lattice point. The lattice constant is the average distance between neighbouring nodes. From these positions the distance $d_{ij}$ between connected nodes $i$ and $j$ is calculated for all $i$ and $j$. Further the average radius of the tubes are chosen by random in the interval $(0.1d,0.1d+0.3d_{ij})$, where $d$ is the lattice constant. This means that there is a minimum radius assuring that no tubes are effectively blocked. Further the maximum radius is related to the length of the tubes in such a way that short tubes cannot be unreasonably wide. The question of the importance of geometry and topology is addressed in the discussion in section \nolinebreak \ref{sec:discussion}.

We consider two fluids within this system of tubes. In Figure \ref{fig:networks} the nonwetting fluid is shown in black, and the wetting fluid is shown in gray. They are separated by a set of interfaces, menisci, in the tubes. We do not allow for film flow. Motion of the fluid during a simulation is represented by the motion of the menisci. In each tube we allow zero, one or two menisci. If at any instant the evolution of the system generates a third meniscus in one tube, the three menisci are collapsed into one. The position of this new meniscus is the one which preserves the volumes of the two fluids. This upper limit of two menisci in each tube sets the resolution of the fluid distribution.

With respect to permeability the tubes are treated as cylindrical, but with respect to capillary pressure they are hour-glass shaped. This means that over each of the menisci in the system there is a capillary pressure which varies with the menisci's positions in the tubes. The formula for the capillary pressure is
\begin{equation}
 p_{\rm c} = \frac{2\gamma}{r} (1-\cos{(2\pi x)}) ,
\label{eq:capilpres}
\end{equation}
which is a modified form of the Young-Laplace\footnote{Young-Laplace law is sometimes referred to as Laplace's equation({\it e.g.} \inlinecite{D92}).} law \cite{D92,AMHB98}. Here, $r$ is the radius of the tube, and $\gamma$ is the inter-facial tension between the fluids. Further, $x$ is a position variable which runs between zero and one. This functional form is mapped onto the central 80 percent of each tube. The closest 10 percent to the nodes is a zero capillary pressure zone. We have done numerical experiments which show that the results from the simulations are insensitive to the exact choice of 10 percent. In particular, we have checked that 6, 8, 10, 12, and 14 percent all work equally well. The major purpose of the zero capillary pressure zone is to enhance the numerical performance of the model. Numerical and unphysical oscillations are prevented, and this makes it possible to run the simulations with larger time steps.

\section{Flow Conditions and Boundary Conditions}
\label{sec:flow}

Before we turn to the discussion of the globally applied pressure and the boundary conditions, we consider the flow equation for a single tube. The volume flux through one single tube is given by the Washburn equation for capillary flow \cite{W21};
\begin{equation}
\label{eq:wash}
  q = - \frac{k}{\mu_{\rm eff}}\cdot\frac{\pi r^2}{d} \left( \Delta p - p_{\rm c}\right) .
\end{equation}
The tubes are considered as cylindrical with radius $r$. The fraction $\pi r^2 / d$ is the cross-sectional area divided by the length of the tube. The permeability of the tube $k=r^2/8$ as is known from Hagen-Poiseulle flow. When two fluids are present in a tube, an effective viscosity $\mu_{\rm eff}$ is used. The viscosities of the two fluids are weighted according to their volume fraction at the beginning of the time step to give $\mu_{\rm eff}$. Further, the pressure $p$ is a discretized field variable with a value in each node. The flow in a tube is proportional to the difference in pressure, $\Delta p$, between the nodes at the ends of the tube. In addition, the sum of capillary pressures $p_{\rm c}$ of the menisci in the tube is subtracted from the pressure difference to give Equation (\ref{eq:wash}) \cite{W21}.

The traditional use of network simulators has been the study of invasion processes. Usually the first row of nodes is an inlet row where all the nodes have the same fixed pressure. Similarly the last row works as an outlet row, typically having all nodes fixed at zero pressure. The point being that for a given time step the pressures in these two rows are fixed constraints. The pressures in all the other nodes are free variables that are solved for. The equation that one needs to solve is the conservation of fluid flux passing through each node. This is nothing else than a discrete Poisson problem for pressure.

Once the pressure field is known, Equation (\ref{eq:wash}) gives the flux in each tube. Many authors have studied invasion under a constant applied pressure. The mother model has been used to study constant flux invasion. By solving the flow field for two different globally applied pressures giving two different fluxes, one can calculate the pressure that would give the desired flux. How to do this in practice is described in detail by \inlinecite{AMHB98}. We will just remark for now that all simulations presented in this paper are done with constant flux rate.

Computer power is limited, and in general a detailed model implies small networks, and large networks imply coarse models. Our model aims at resolving the dynamics due to variable capillary pressure when passing through throats. It also aims at having a realistic distribution of the two fluids within the pore space as this has consequences for the transport properties of the system. Roughly a system of $100\times200$ nodes is the maximum size available for simulations with our model within reasonable time on single CPU computers. When many data points are needed to probe through parameter space, $40\times 80$ nodes is more convenient.

\subsection{Biperiodic Boundary Conditions}
Invasion simulations go on until the invading fluid reaches the outlet. If they were to go on further, they would change character since one fluid is percolating the system. We wish to address the question of what happens in a system far from inlets and outlets. That is, given a very large system, we take out a small piece somewhere in the middle and study its properties. 

This is done by adjoining the inlet row and outlet row so that the fluid that flows out of the last row enters the first row. In practice this works in such a way that the simulation can go on for ever, regardless whether one fluid percolates the system or not. The time evolution in Figure \ref{fig:networks} illustrates how the fluid, which leaves the top row, enters the bottom row. In a sense having biperiodic boundary conditions make the system infinite. However, the system is closed, and there is a fixed volume of each fluid in the system. Thus, each simulation takes place at constant saturation equal to the one of the starting configuration. We return to this point when presenting results from simulations in section \nolinebreak \ref{sec:simulations}.

The straightforward appliance of a global pressure between the inlet row and the outlet row in invasion simulations needs modification. One could do this by adding one row of ghost nodes on both sides of the system. The pressure would be fixed on these two rows as before. The equations would be solved as before, only with the additional constraint that the outgoing flux in each tube between the last row and its ghost row should equal the incoming flux in the respective tube between the other ghost row and the first row. This method works, but has two drawbacks. One is that it gives more complicated computer code. The other is the fact that the pressure is still fixed to have the same value along a straight line through the system, thus giving a boundary effect.

Therefore instead of using ghost nodes with fixed pressures as the driving force, we make a jump in the pressure over all the tubes on the boundary. This has been implemented in random resistor networks \cite{roux,BH98}. It works so that whenever looking across the boundary from one side one sees a positive pressure jump. Whenever looking from the other side one sees a negative jump. The value of the jump is the same in all tubes along the boundary, and will be referred to as the global pressure. The global pressure is the driving force making the fluid go around the system. In order to visualize these boundary conditions one can think of the network as placed on the surface of a torus. The flow is around the torus, but one cannot see any trace of the adjoint boundary. That is the advantage of the pressure jump technique. Otherwise the two techniques give similar results.

\section{Updating the Menisci}
\label{sec:menisci}
In the model the entire volume of the porous medium is contained within the tubes. Further, all interfaces between the two liquids are also situated within the tubes. When the fluids flow, all interfaces move according to the flow field. This is straightforward within a single tube. In nodes, however, one needs to define a set of moving rules based on physical reasoning. We also provide an algorithm, which purpose is to generate more realistic motion of two menisci when they are simultaneously approaching the same node.

\subsection{Updating the Menisci Within the Tubes}
The solution of the flow equations provides information on the flux at each instant in every tube. Knowing the cross-sectional area of every tube, we calculate the velocity in every tube from the fluxes. The evolution in time is simply given by explicit Euler integration; every meniscus is moved a distance equal to the velocity in the respective tube multiplied by the time step. The time step is chosen so that no meniscus should move a longer distance than, say, ten percent of a tube length. Most of the menisci will remain within the bounds of the tubes after the time step. However, some of the menisci have been moved out of the tubes and must be dealt with.

\subsection{Moving the Menisci Across the Nodes}
The procedure of moving menisci across nodes needs some elaboration. In the mother model \cite{AMHB98} this was done in a way which was not conserving the volumes of the fluids. This was not a problem in the mother model, as the cumulative volume loss over the relevant time scales was small. However, in the present case the model is run considerably longer, and whence these volume losses are no longer acceptable. Also, when simulating an invasion process these small volume changes are easy to come around with the following trick. The invading fluid is coming from an inlet row and the defending fluid is running out on the outlet row, both at a constant flow rate. A change in volume is therefore equal to a change in elapsed time.

\begin{figure}
\centering
\includegraphics[height=8cm]{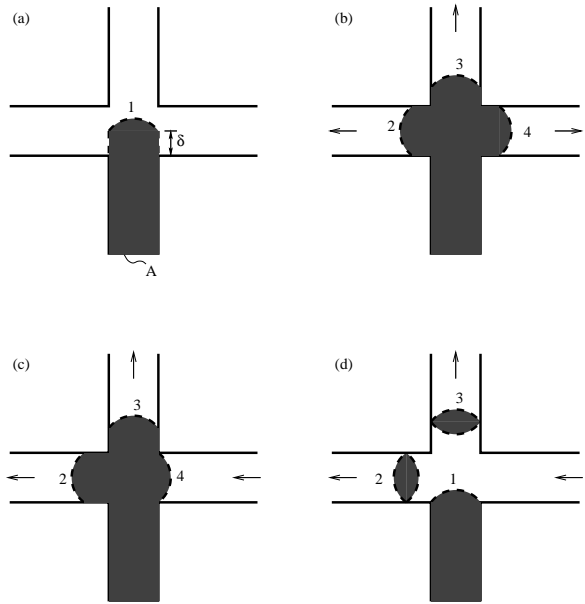}
\caption{(a)\textsl{The shaded fluid is entering the node from one tube. The meniscus {\it no.}1 is moved a distance $\delta$ into the node, which corresponds to an incoming volume; $\delta A$.}
(b)\textsl{The case of three tubes carrying outgoing flux. The arrows indicate the flow direction.}
(c)\textsl{The case of two tubes carrying outgoing flux. Note that in this case the meniscus {\it no.} 4 is left at the edge of the right tube.}
(d)\textsl{As} (c)\textsl{one time step later.}
}
\label{fig:onemen}
\end{figure}

We want to simulate a closed system with fixed saturation, {\it i.e.}, having volume conservation. The strategy is to look at each node and find the menisci which have passed the boundaries of the tubes and are actually `moved into' the node. In the simplest case, only one meniscus is inside the node as shown in Figure \ref{fig:onemen}(a). The meniscus has been moved a distance $\delta$ into the node. This distance times the cross-sectional area of the tube is equal to an amount of volume, $V=A\delta$, which must be moved over to the neighboring tubes. Again in the easiest case the fluid in all three neighboring tubes is flowing away from the node. Then we simply state that we should create one meniscus in each of the three tubes. They should be placed at distances from the node so that i) the three volumes add up to the one having flown into the node, and ii) the ratios between these three volumes are the same as the ratios between the fluxes in the same three tubes. We illustrate this case in Figure \ref{fig:onemen}(b).

This rule does not hold in the case when in one, or more, of the other tubes the fluid flows towards the node instead of away from the node. Since we consider immiscible flow we cannot allow the fluids to mix, though somehow that would be the natural idea when two different fluids meet in a node. Within a single time step we allow for a small `error'. We place a meniscus on the edge of the tube in which the fluid flow is towards the node. See Figure \ref{fig:onemen}(c). The menisci in the other two tubes are placed according to the same principles as before, only now applied to two tubes instead of three. Physically, this is equivalent to saying that all of the white fluid coming from the right on the figure pass the node before the shaded fluid enters the node. The roles of the fluids change in the next time step giving the situation shown in Figure \ref{fig:onemen}(d). We observe how bubbles have been created in the two tubes carrying outgoing flux. This is an effective mixing of the fluids. This mechanism is the same as in the mother model, and we refer to the discussion there.

\begin{figure}
\centering
\includegraphics[height=4cm]{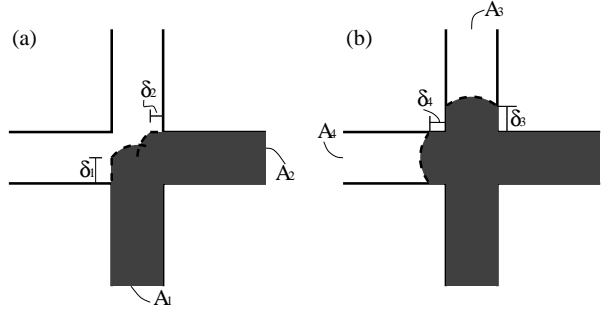}
\caption{(a)\textsl{Two menisci are moved into the node in the same time step.}
(b)\textsl{Two tubes carrying outgoing flux. The menisci positions $\delta_3$ and $\delta_4$ are chosen so that outgoing volume is equal to incoming volume, and so that the ratio $\delta_3/\delta_4$ is equal to the flux ratio in the two tubes.}
}
\label{fig:twomen}
\end{figure}

These rules generalize quite straightforward to the other possible cases where two or more menisci have entered the node within the same time step. Consider the situation in Figure \ref{fig:twomen}(a). The two menisci within the node represent two inflowing volumes. We add the volumes together
\begin{equation}
  V_{in}=A_1 \delta_1 + A_2 \delta_2 .
\label{eq:vol1}
\end{equation}
Menisci are created in the other two tubes according to the same principles as before. See Figure \ref{fig:twomen}(b). This ensures volume conservation,
\begin{equation}
  V_{in} = V_{out}=A_3 \delta_3 + A_4 \delta_4 .
\label{eq:vol2}
\end{equation}

Also here it may happen that the fluid flow in one tube is towards the node. This is treated in the same manner as before, implying that the last tube will carry all the outgoing flux. These rules work very well. The only scenario not described so far is in the case of having very small bubbles in a tube. During a time step both of the menisci enclosing the bubble may be moved into the node. Then, one simply has to split the moving job into two parts. First, one ignores the second meniscus, or menisci if bubbles are coming from more than one tube, and updates everything as before. Then one repeats the procedure considering the second meniscus, or menisci, according to exactly the same rules.

We remark here that when two menisci enter the node within the same time step, they are joined together in the sense that no bubbles are created. If they were to arrive to the node closely after each other, bubbles of the withdrawing liquid will remain in the tubes. Consequences of this will be addressed in the next subsection.

\subsection{Imitating the Merging of Menisci in the Nodes}

\begin{figure}
\centering
\includegraphics[height=8cm]{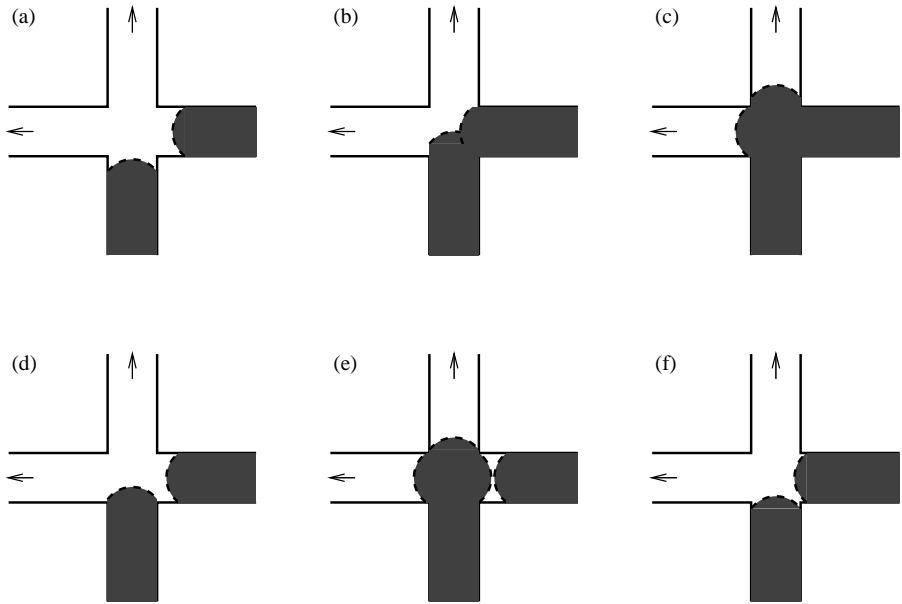}
\caption{\textsl{The detailed evolution around a node is dependent on the size of the time step.}
(a)\textsl{Two menisci are approaching the same node from neighbouring tubes.}
(b)-(c)\textsl{The time step is just long enough such that both menisci enter the node within the time step. The menisci are merged - there are no bubbles created.}
(d)-(e)\textsl{The time step is a little shorter and only one meniscus enters the node. A bubble is formed in the tube to the right, and the menisci are not allowed to join.}
(f)\textsl{Adjustment of the two menisci in }(d)\textsl{ such that they will both enter the node in the next time step.}
}
\label{fig:menmer}
\end{figure}

In our time integration we need to choose a time step. We do not have a fixed time step, but choose one that does not move any meniscus more than a certain fraction of the tube length. The velocity field in the network will contain local velocities that differ by several orders of magnitude, due to geometrical heterogeneity of the network or due to the current fluid configuration. Local velocities and the choice of time step imply that sometimes or some places two menisci will arrive at the same node within the same time step. We illustrate this situation in Figure \ref{fig:menmer}. Here, we see that the two menisci in Figure \ref{fig:menmer}(a) could reach the node in the same time step, which gives the evolution shown in the Figures \ref{fig:menmer}(b) and \ref{fig:menmer}(c). If the time step is shorter, so that only one of the menisci reaches the node during the time step, the situation will be like in the Figures \ref{fig:menmer}(d) and \ref{fig:menmer}(e).

The time step is often recognized as one of the parameters of a dynamical system. The fact that the exact evolution of the system will depend on the choice of time step is unavoidable. In our case, this does not really matter as it is the statistical properties that we seek, and they should not change. However, a large time step will in general allow for many pairs of menisci to meet in the node during the time step, and whence be joined together. A small time step will in general make one of the menisci arrive earlier than the other in otherwise same situations. These menisci are then not joined together. This time step dependency is clearly unphysical, and we wish to reduce it as much as possible.

This procedure is actually quite important because we know that an actual porous medium has volume in its pores. It takes some time to fill a pore, and menisci will not just pop over to the other side instantaneously. Here, there is a difference between a drainage step and an imbibition step. In drainage when a meniscus passes through the threshold of a throat entering into the pore, there has been a local pressure buildup giving rise to a small burst \cite{H30,LZ83,MFFJ92}. The pressure buildup and subsequent release make the flow into the pore quite fast, and the time to fill the pore quite short. In imbibition the capillary forces make the pores the hardest areas to pass, meaning that wetting fluid needs more time to fill up the volume of the pores.

In our model menisci are moved instantaneously across pores. This is unphysical. If the physical situation is that two menisci enter into the pore volume at the same time, then some kind of mixture is likely to happen. Bubbles can be created or the menisci can join together pushing the other liquid out on the other side first. The circular disk model, successfully studied by \inlinecite{CR90}, provides some basic intuition on this problem. We are dealing with a situation that globally is neither wetting nor nonwetting invasion, due to the biperiodic topology of the network. Locally, both wetting and nonwetting invasion steps happen all the time. The detailed knowledge about how menisci advance through pore volumes \cite{CR90} is too complicated to be incorporated fully in our model.

Instead, our model aims at grasping the bare cut essence of adjoining menisci in pores. First we associate an effective volume to each node. Since the node is the meeting point of four tubes, it is reasonable to assume that the volume in the end of each tube add up to an effective node volume. We have chosen to say that 20\% of each tubevolume is added to the respective pore volumes. The curvature of the tube when modeling capillary pressure starts at a distance that is 10\% of the tubelength away from the node, see Equation (\ref{eq:capilpres}). 20\% is roughly where the tubes start to get narrow. We have done tests which show that the results are not very sensitive to the excact choice of $20\%$.

Thus, having assigned an effective volume to every node, we can estimate the time needed to fill the nodes whenever a meniscus has entered into a node. According to the position of the meniscus in the node we know how much volume that is already in the node. Here, one should include the 20\% endpart of the tube from which the meniscus arrived. Knowing the instantaneous flux in the tube from which the meniscus came, we can calculate an estimate for the filling time by taking the remaining volume and divide it by the flux. Further, since we know the velocities in the other tubes, we can for every meniscus that is moving towards the node calculate an estimate for the arrival time. If this arrival time is shorter than the filling time for the meniscus that has arrived to the node, one can say that physically these two menisci should have been present in the node at the same time.

In our model, we therefore assume that the menisci physically would have joined together, {\it i.e.}, arrived into the node in the same time step. The time evolution is organized as follows. First all menisci are moved after the naive Euler scheme. Then for every node we check if there is one single meniscus inside it. If it is, we calculate its filling time and find the shortest arrival time for any other meniscus to that node. If these times show that the menisci physically should have met in the node, we move both menisci so that they are equally distant from the node, see Figure \ref{fig:menmer}(f). This is done under the constraint of volume conservation. Then it might happen that after this movement both menisci are inside the node or they are both outside the node. In either case the procedure has now come to moving the menisci which are still inside the nodes. This is done as described in detail in the previous subsection. Eventually, the flow equations are solved anew for the new fluid configuration and the whole process is repeated.

In this subsection we have described how we simulate merging of menisci in nodes. The mother model did not have this feature, but it was used to study drainage invasion processes. In an invasion process the number of menisci arriving at a node approximately at the same time is quite small. Here, we have biperiodic boundary conditions allowing for simulations to go on for a much longer period of time. In particular, it allows the system to reach fully developed flow far from initial fronts and configurations. Fully developed two-phase flow is reported on in Avraam and Payatakes (\citeyear{AP95a,AP95b,AP99}), where the number of bubbles or clusters of the fluids is large. Therefore, it is important to take into account not only the possibility of breaking up bubbles, which happens whenever passing through a node, but also this mechanism of merging bubbles based upon physical reasoning.

In practice during a simulation this routine makes adjustments so that both nonwetting and wetting pairs of menisci are joined. However, this happens more often in the wetting case. The routine has a larger effect of making the wetting phase more compact than for the nonwetting phase. This is in agreement with the well-known fact that considerable smoothing of the invasion front takes place in wetting invasion \cite{LZ83,CR90}.

\section{Simulations}
\label{sec:simulations}
We will in this section provide simulation results. The general nature of the simulations will be addressed in the first subsection. Thereafter, we discuss the relevance and importance of possible parameters of the system. They include viscosities, interfacial tension, flow rate, and system size. Finally, detailed results are given for the case of two fluids with equal viscosity using the capillary number as parameter.

\subsection{The Nature of the Simulations}

We start out by giving data samples from two simulations before turning to generalities. The system is $20 \times 40$ nodes in both simulations. The initial configuration of the systems is one horizontal region with wetting fluid and another with nonwetting fluid. In the first sample the nonwetting volume fraction, the saturation, is 52\%. In the second case it is 42\%.

Both fluids have the same viscosity; $\mu = 1.0\ {\rm P}$. The interfacial tension between the fluids is $\gamma = 30.0\ {\rm dyn/cm}$. The flow rates are $Q_{\rm tot} = 1.38 \times 10^{-3}\ {\rm cm}^3/{\rm s}$ and $Q_{\rm tot} = 1.38 \times 10^{-2}\ {\rm cm}^3/{\rm s}$, respectively. The capillary number is defined as follows
\begin{equation}
  C_{\rm a} = \frac{\mu Q_{\rm tot}}{\gamma \Sigma},
\label{eq:capilnum}
\end{equation}
where $\Sigma$ is the cross-sectional area of the network. Physically the capillary number is the ratio between viscous and capillary forces. Our two cases have $C_{\rm a} = 3.2 \times 10^{-4}$ and $C_{\rm a} = 3.2 \times 10^{-3}$.

The total flux $Q_{\rm tot}$ is equal to the sum of the wetting flux and the nonwetting flux. During the simulations the total flux is held at a fixed value. However, it is the physics of the systems that determine the flux of the wetting and of the nonwetting fluids, respectively. Fluid motion means reorganization of menisci and capillary forces change. In general, when the viscosities of the fluids are different, the effective viscosity of the fluid in each tube also change along with the time evolution. In order to keep the total flow rate constant, the globally applied pressure must be adjusted in every time step.

\begin{figure}
\centering
\begin{tabular}{cc}
\includegraphics[height=4.8cm]{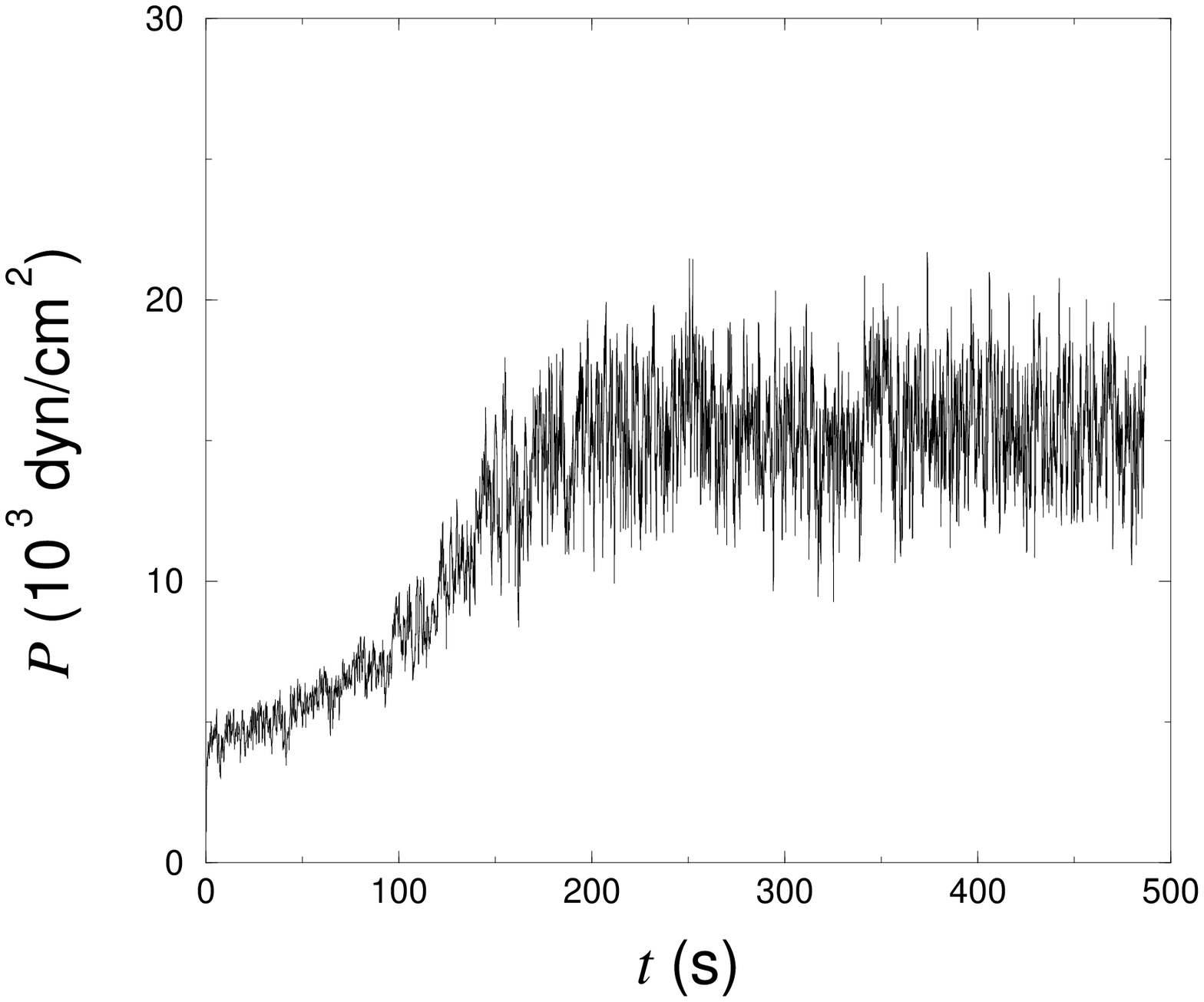} &
\includegraphics[height=4.8cm]{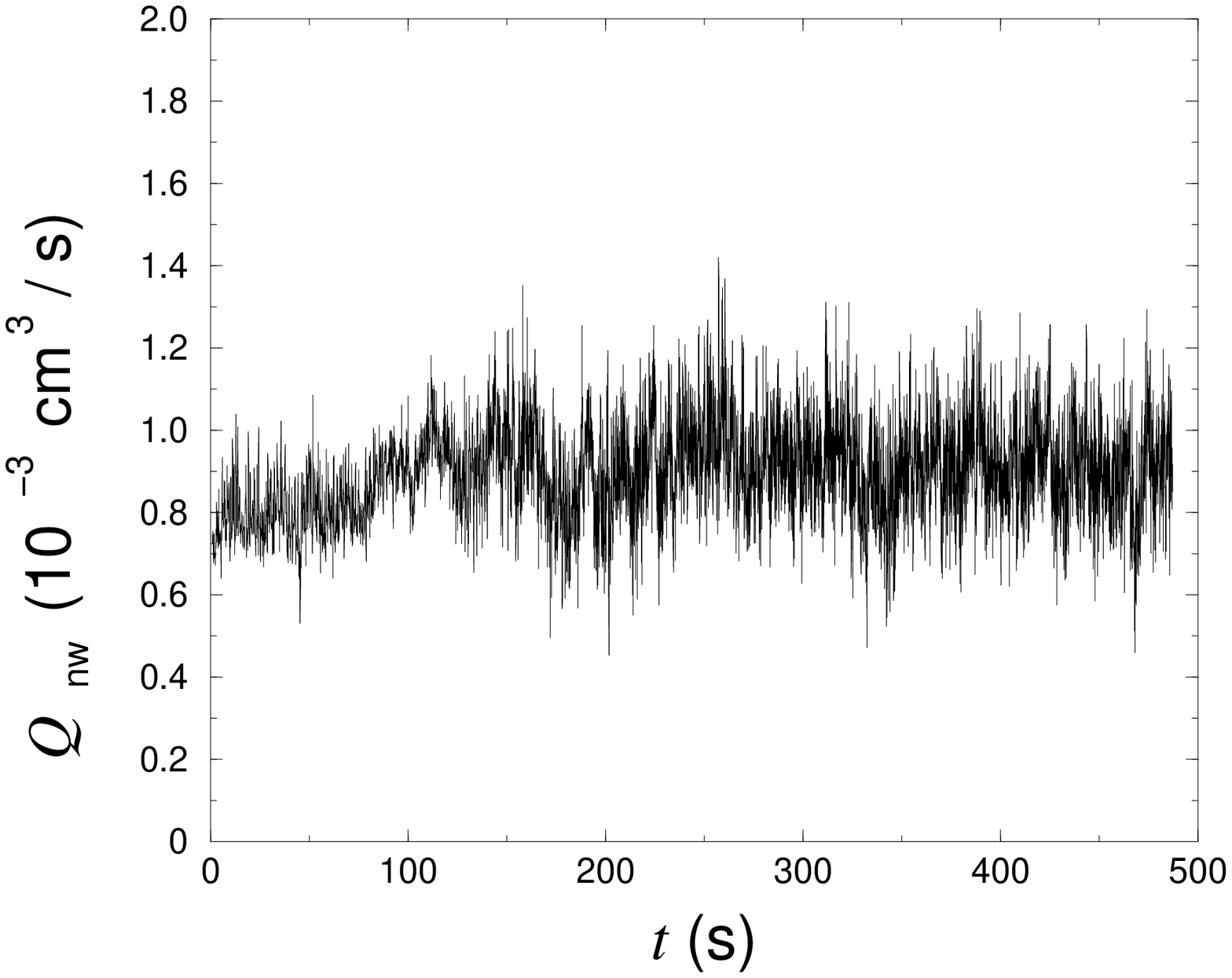}
\\ (a) Global pressure & (b) Nonwetting flux
\end{tabular}
\caption{\textsl{Typical time evolution of a simulation. Here the nonwetting saturation is about 52\% and the capillary number is $C_{\rm a}=3.2 \times 10^{-4}$. The fluid travels on the average approximately twice around the system during this simulation.}
}
\label{fig:flux1}
\end{figure}

\begin{figure}
\centering
\begin{tabular}{cc}
\includegraphics[height=4.8cm]{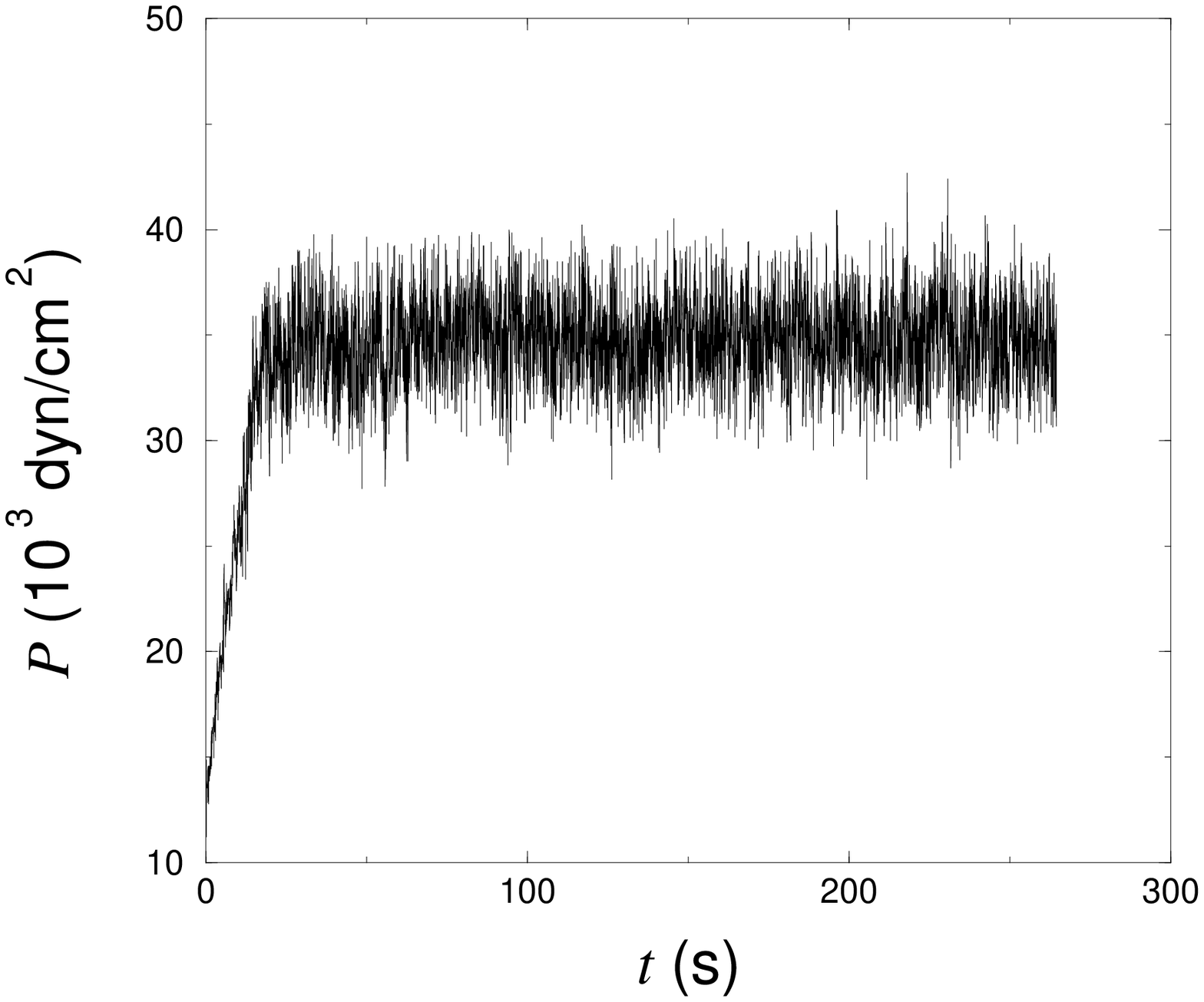} &
\includegraphics[height=4.8cm]{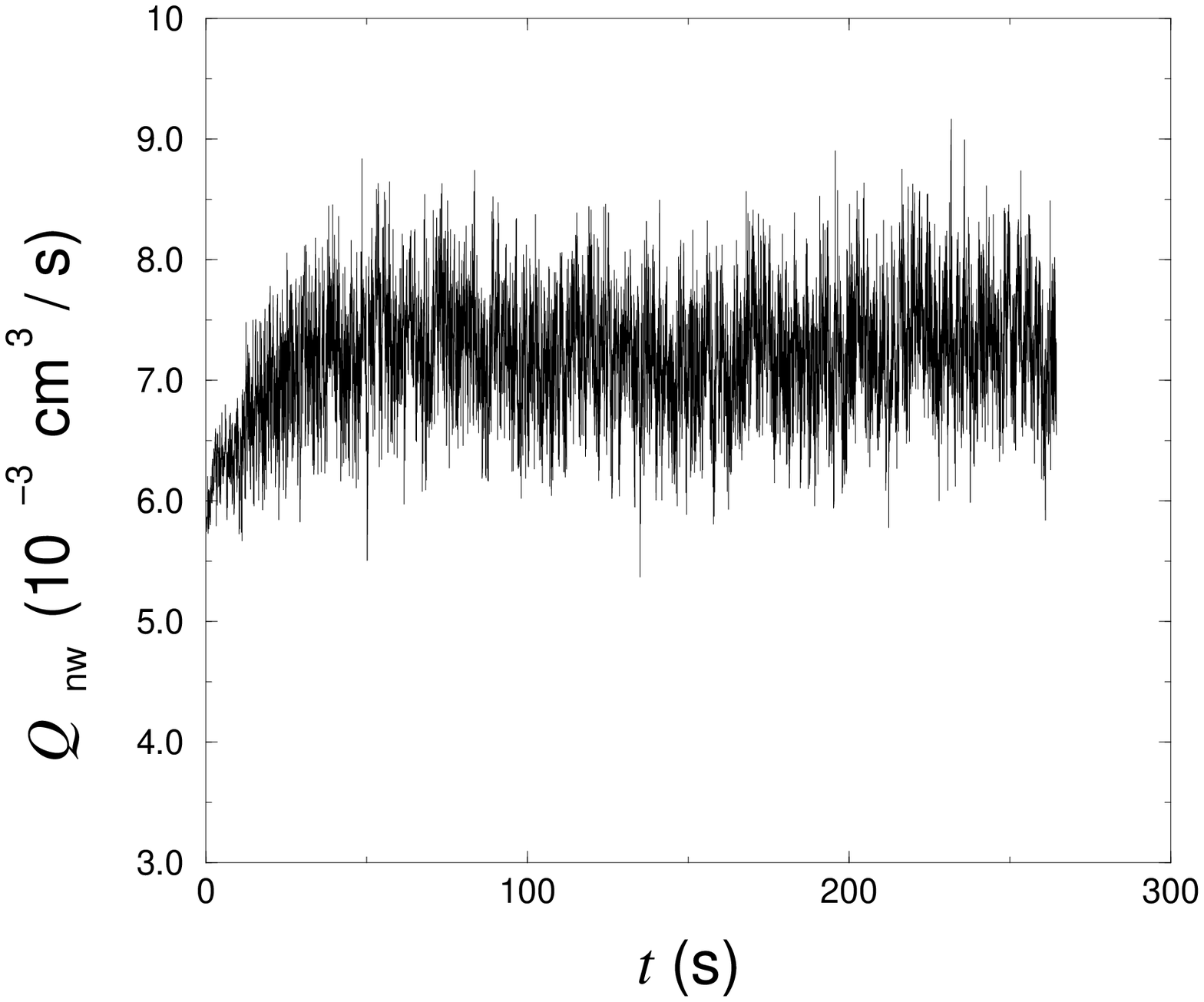}
\\ (a) Global pressure & (b) Nonwetting flux
\end{tabular}
\caption{\textsl{Typical time evolution of a simulation. Here the nonwetting saturation is about 42\% and the capillary number is $C_{\rm a}=3.2 \times 10^{-3}$. The fluid travels on the average approximately eight times around the system during this simulation.}
}
\label{fig:flux2}
\end{figure}

The results from the two samples are shown in the Figures \ref{fig:flux1} and \ref{fig:flux2}. They show the global pressure and the nonwetting flux versus time. The curves are noisy on the time scale used in these plots. One should note that the time axis is compressed so that the data points are close. What looks like a jump on the plot may very well be 20 or 100 time steps in the simulation.

The curves for the global pressure are very characteristic. There is a transient part and a steady part. This means that at the beginning of a simulation, the system remembers its initial configuration. After some time, which depends on system parameters, the system reaches a steady-state. In Figure \ref{fig:flux1}(a) the capillary number is a factor ten smaller than in Figure \ref{fig:flux2}(a). We can see how this makes the transient period longer in the first case.

The initial configuration, transient part and steady part of a typical system, $C_{\rm a}=1.0\times 10^{-3}$ and $S_{\rm nw}=50.2\%$, are illustrated in Figure \ref{fig:networks}. Here snap-shots of the fluid distribution are shown at different times. We are not really interested in the transient part. The important aspect is that the systems reach a steady-state. In this state all fluid is mobilised. We know this by looking at consecutive images of the fluid distribution. This means that the steady-state is characteristic of the medium and the fluids; the state will be reached whatever initial configuration. We will return to possible history dependence for other capillary numbers than depicted here in the next subsection.

Figures \ref{fig:flux1}(b) and \ref{fig:flux2}(b) show the nonwetting flux averaged over the entire system as a function of time. The curves are not normalized with respect to total flux. We merely wish to show their character. Their transient parts are not so explicitly distinct from their steady parts as is the case for the global pressure evolution in these samples. In general, we have used both curves to determine a time after which the systems for sure have reached the steady-state. When the system has reached the steady-state we have found the time average of the global pressure, the wetting flux, $Q_{\rm w}$, and the nonwetting flux, $Q_{\rm nw}$. Since the simulations have gone on for in the order of 50000 time steps after reaching this state, the statistics are good. We focus the rest of the presentation of results on one of the more interesting properties of the systems. That is, the fractional flow of one of the fluids as a function of the system parameters.

\subsection{Relevant Parameters}

\begin{figure}
\centering
\includegraphics[height=6cm]{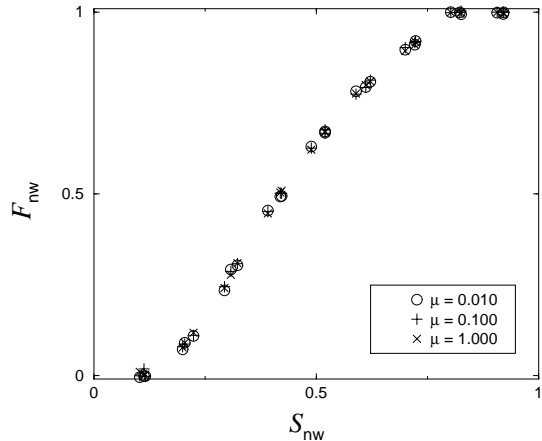}
\caption{\textsl{The nonwetting fractional flow versus the nonwetting saturation for capillary number $C_{\rm a}=1.0\times 10^{-3}$. The system size is $20 \times 40$. The viscosity of both fluids have been given the same three different values. We observe the data-collapse which is expected.}
}
\label{fig:visc}
\end{figure}

\begin{figure}
\centering
\includegraphics[height=6cm]{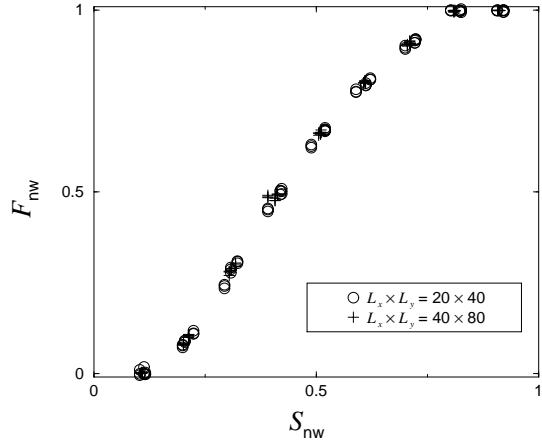}
\caption{\textsl{The nonwetting fractional flow versus the nonwetting saturation. Two different system sizes, $20\times 40\ $ and $40\times80$, have been used in the simulations. This data collapse shows the size independent nature of the simulations.}
}
\label{fig:size}
\end{figure}

One important parameter of the system in these simulations is the nonwetting saturation, $S_{\rm nw}$, which is the volume fraction of the nonwetting fluid. The corresponding output is the fractional flow of the nonwetting fluid
\begin{equation}
  F_{\rm nw} = \frac{Q_{\rm nw}}{Q_{\rm tot}}.
\end{equation}
Figures \ref{fig:visc} and \ref{fig:size} show the nonwetting fractional flow as a function of the nonwetting saturation. The S-shape is resemblant of the one found in Buckley-Leverett fractional flow \cite{D92}. Both figures show the curves obtained for $C_{\rm a}=1.0\times 10^{-3}$. The shapes and their dependency of $C_{\rm a}$ will be discussed in the next subsection.

We have done simulations for fluids with equal viscosity. The interfacial tension has been $\gamma=30.0\ {\rm dyn/cm}$ for all simulations. The total flux has been changed from simulation to simulation in order to give the desired capillary number after Equation (\ref{eq:capilnum}). We wish to demonstrate that for the case of equal viscosities, the capillary number is the relevant parameter for the fractional flow curves.

From Equation (\ref{eq:capilnum}) we see that the capillary number is unchanged if $Q_{\rm tot}$ and $\gamma$ are scaled with the same factor. If $\gamma$ is scaled with a factor, so is all capillary pressures after Equation (\ref{eq:capilpres}). When the scaled $p_{\rm c}$ is inserted into Equation (\ref{eq:wash}) one can simply choose to scale $\Delta p$ and $q$ with the same factor. Cancellation of the factor on both sides of the equation shows that this is the solution of the problem when also the total flux is scaled with this factor. This argument assures that the interfacial tension between the fluids can be chosen to a reasonable value in the simulations. The dependence on the capillary number will be the same.

Keeping the surface tension $\gamma$ fixed, and thus $p_{\rm c}$ from Equation (\ref{eq:capilpres}) fixed, this way of reasoning can be used also on the relationship between the viscosity and the total flux. They appear as the product $\mu Q_{\rm tot}$ in Equation (\ref{eq:capilnum}). As long as both fluids have the same viscosity, the effective viscosity $\mu_{\rm eff}$ in Equation (\ref{eq:wash}) is equal to the same $\mu$ for all tubes. If on the other hand the fluids have different viscosity, the effective viscosity will in principle change independently in each tube. So for equal viscosities a change in the viscosity and an inverse change in the flux $q$ will cancel out, in Equation (\ref{eq:wash}), like in the case for the interfacial tension. Figure \ref{fig:visc} shows this effect. Here, the capillary number is fixed at the value $C_{\rm a}=1.0\times 10^{-3}$ while the viscosity is allowed three different values. The system size is $20 \times 40$ nodes and three different random seeds, giving three different geometries, are used. The saturation has been varied between 10\% and 90\%, roughly. The data-collapse is very good as it should be. Three and three data points should according to our reasoning be equal. However, round-off makes the exact evolution of the system different for each viscosity used. Therefore the points will not be identical. The small differences can be considered as a measure of the error bars of the time averaging process.

The last parameter to examine in Equation (\ref{eq:capilnum}) is the cross-sectional area $\Sigma$. The number of nodes $L_x$ in the cross-section is proportional to the cross-sectional area. Further, the number of rows of nodes $L_y$ is also a possible parameter of the curve, even though not included into the expression (\ref{eq:capilnum}). Figure \ref{fig:size} shows the result for two different system sizes; $L_x \times L_y = 20\times 40$ and $L_x \times L_y = 40\times 80$. The capillary number is fixed at the same value as in Figure \ref{fig:visc}. We see that the curves collapse onto each-other. This is a non-trivial result. Typically, invasion simulations give scaling relations between, {\it e.g.}, the front width and the system size. Here we find that the results are size independent. In Figure \ref{fig:size} both $L_x$ and $L_y$ are changed at the same time. One could imagine that there were an individual dependence on each of them that cancels out. We have checked this, and there is not. The size independence must be understood as a result of the construction of the system. The system is forced to have a certain saturation. It is reasonable to think that saturation heterogeneities only are visible below a certain length scale, in terms of number of nodes, within the system. Above this length scale the system has homogeneous saturation. Each system, or part of the system above this scale, should display the same properties, and that is what we see.

\subsection{Fractional Flow Curves}

\begin{figure}
\centering
\begin{tabular}{cc}
\includegraphics[height=5.0cm]{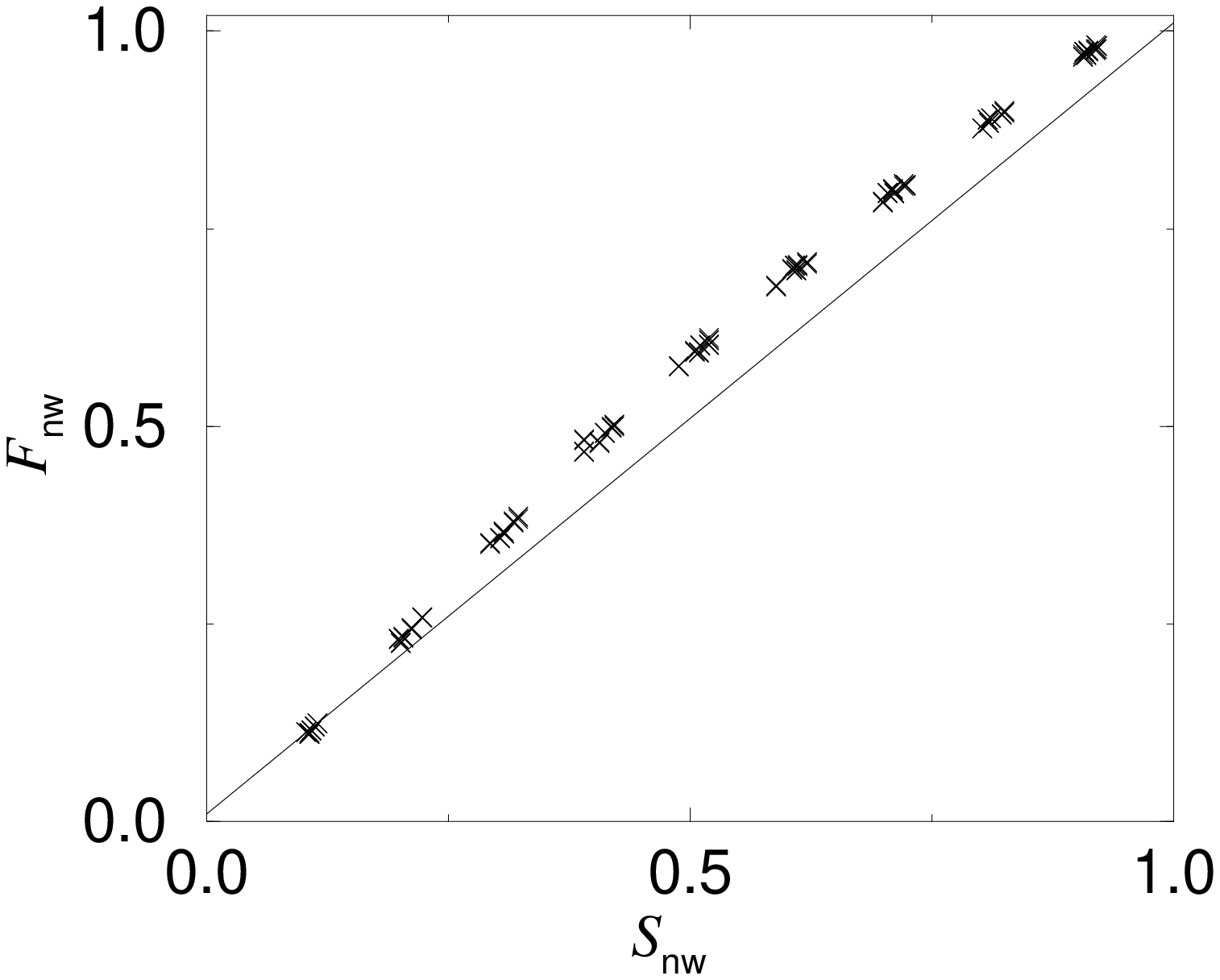} & 
\includegraphics[height=5.0cm]{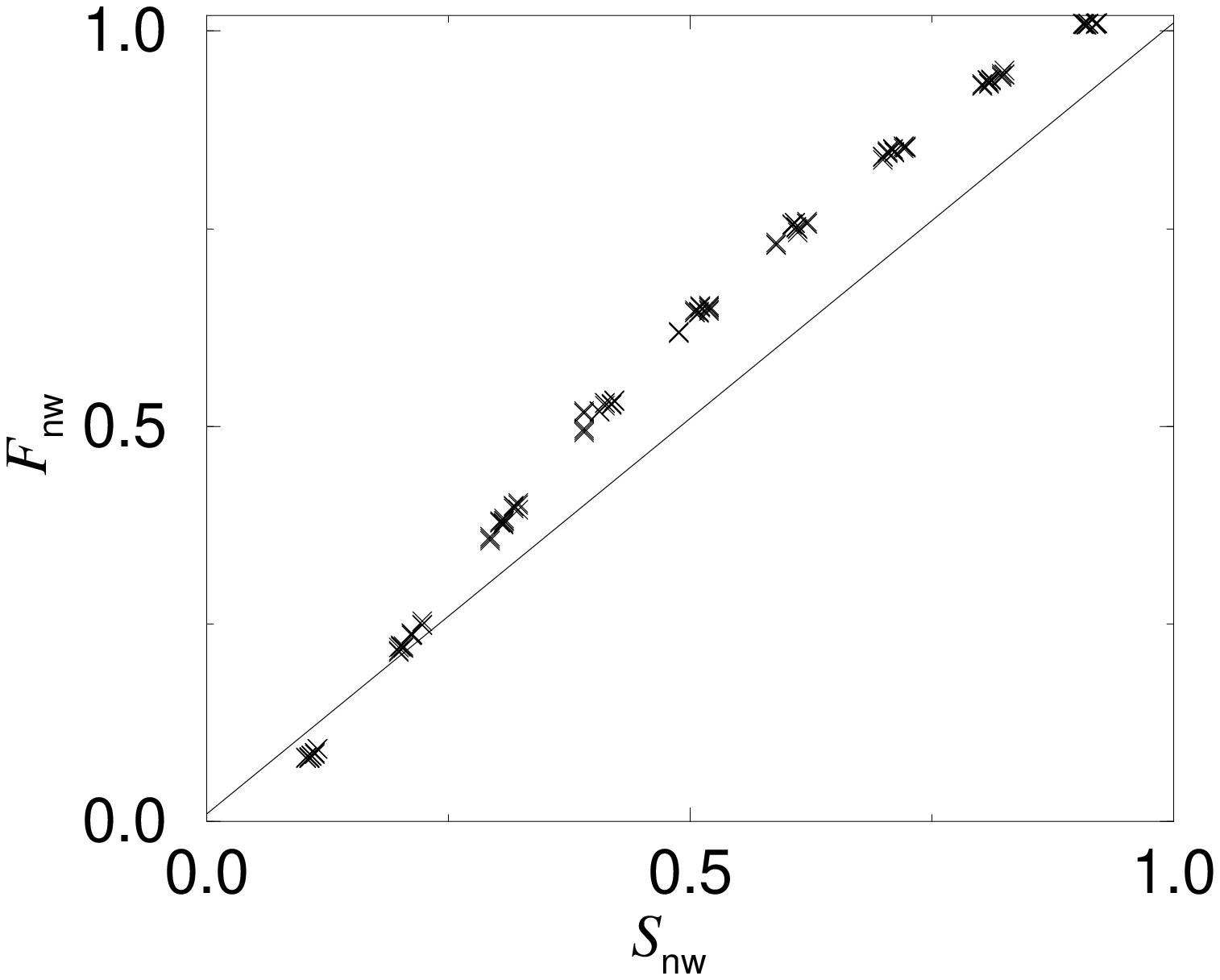} \\
(a) $C_{\rm a} = 3.2\times 10^{-2}$ &
(b) $C_{\rm a} = 1.0\times 10^{-2}$ \\  &  \\
\includegraphics[height=5.0cm]{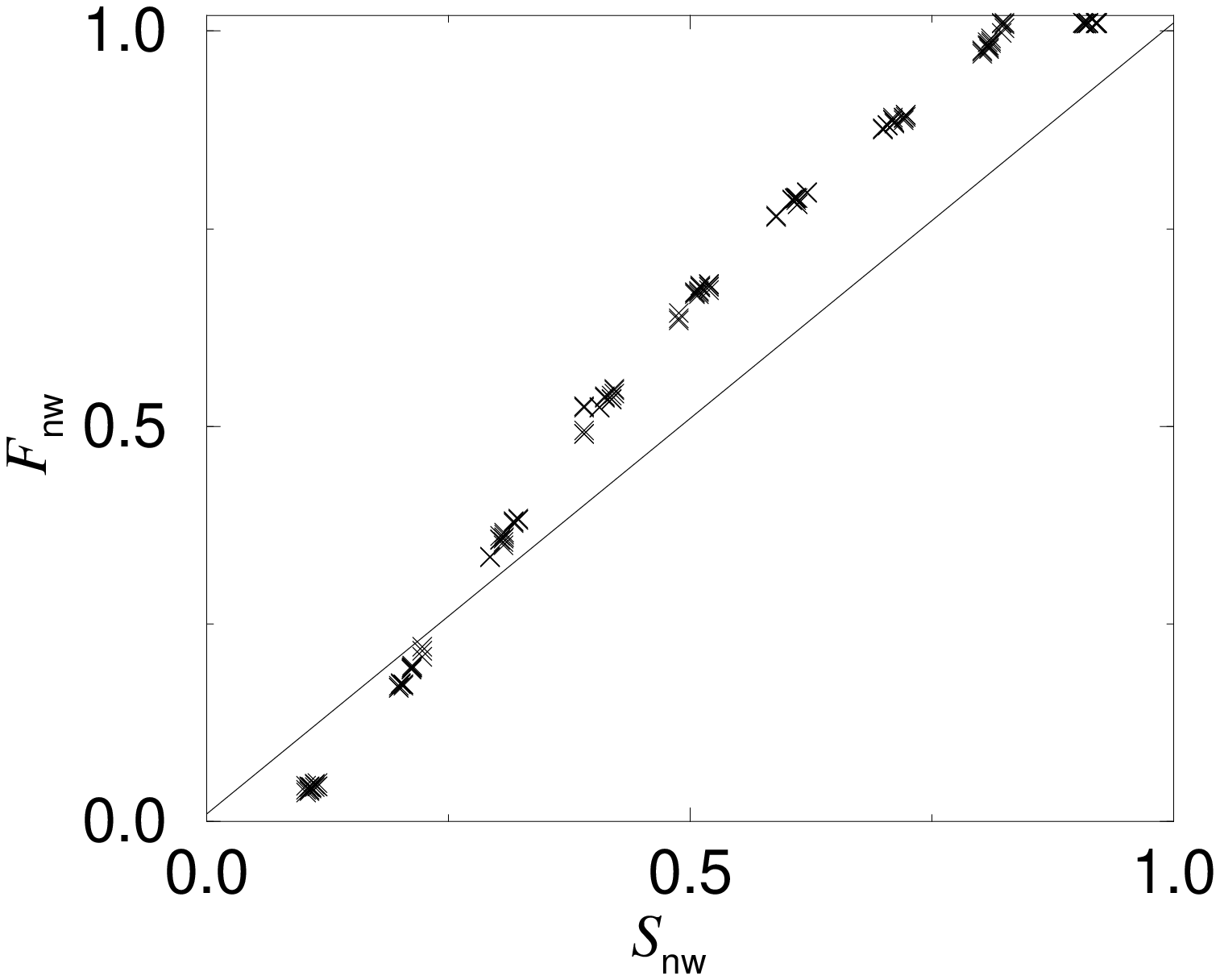} & 
\includegraphics[height=5.0cm]{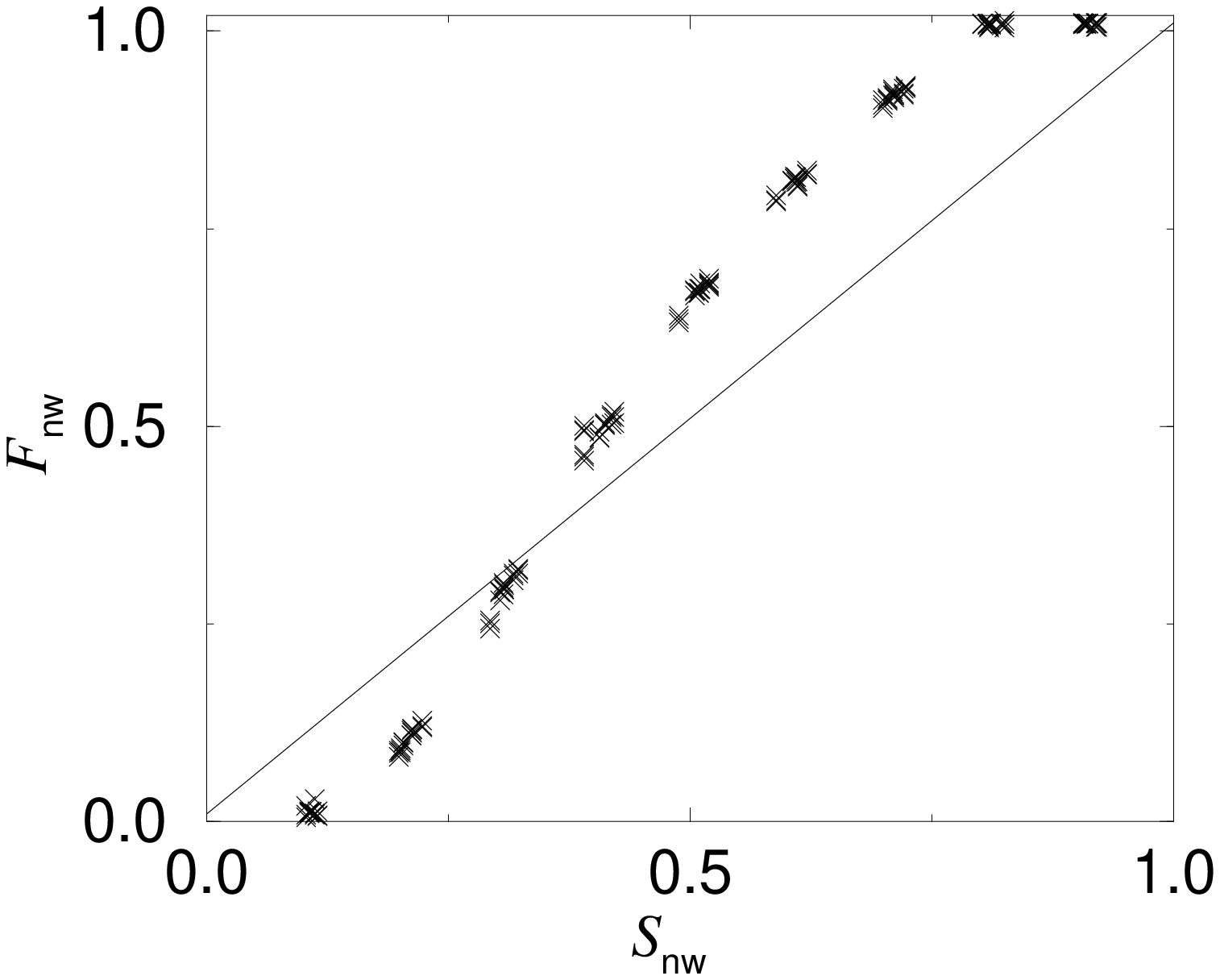} \\
(c) $C_{\rm a} = 3.2\times 10^{-3}$ &
(d) $C_{\rm a} = 1.0\times 10^{-3}$ \\  &  \\
\includegraphics[height=5.0cm]{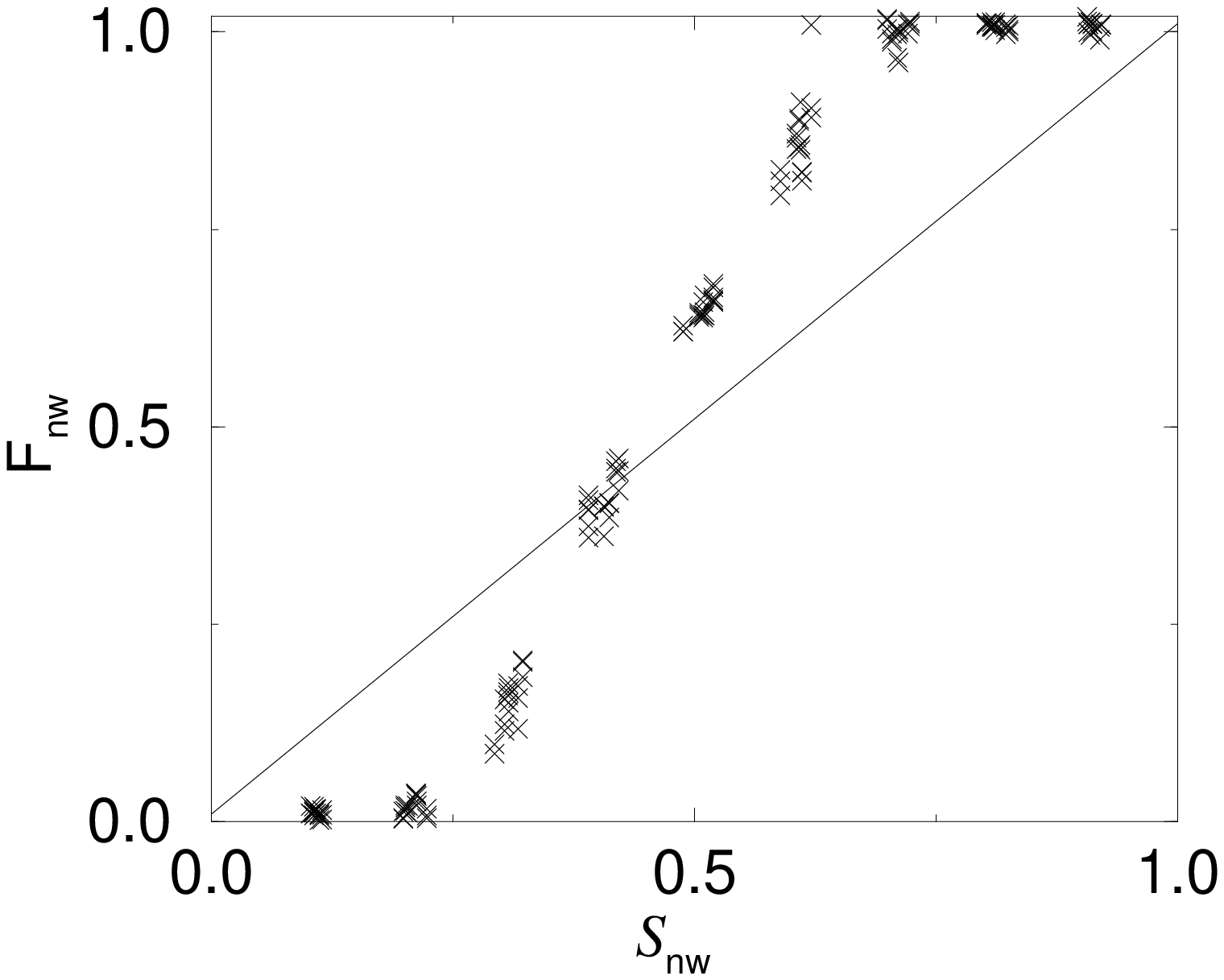} & 
\includegraphics[height=5.0cm]{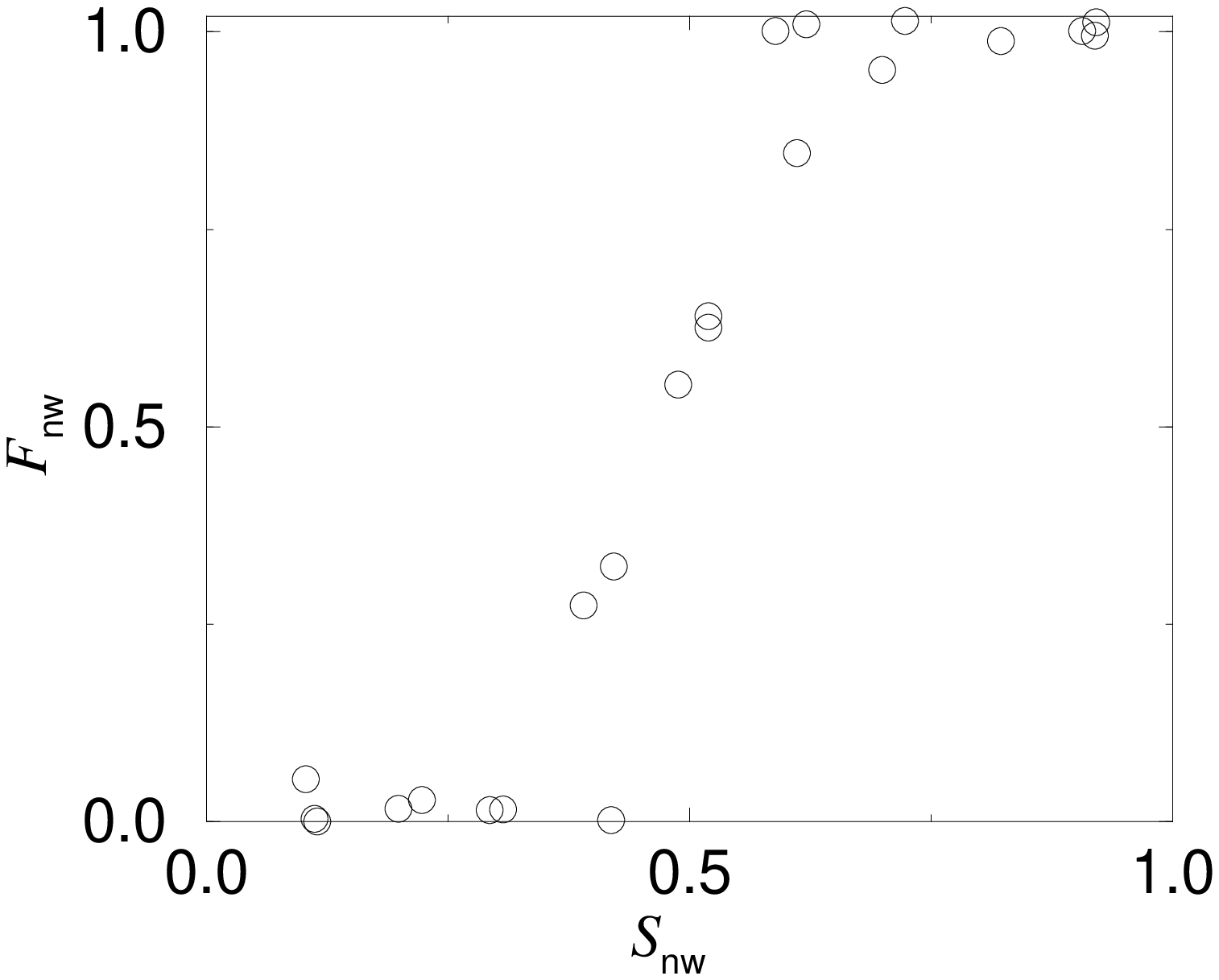} \\
(e) $C_{\rm a} = 3.2\times 10^{-4}$ &
(f) $C_{\rm a} = 1.0\times 10^{-4}$
\end{tabular}
\caption{\textsl{The plots show the nonwetting fractional flow versus the nonwetting saturation for the case of two fluids having equal viscosity. The capillary number is the parameter $C_{\rm a}$.}(a)-(e) \textsl{are in the history independent regime, while } (f) \textsl{is in the history dependent regime and is thus more tentative}.
}
\label{fig:capildep}
\end{figure}

In the previous subsection we found out that for two fluids having equal viscosity, the capillary number serves as the relevant parameter for fractional flow versus saturation curves. Figure \ref{fig:capildep} shows the obtained curves for six different capillary numbers. In (a) to (e) data points for both system sizes $20\times 40$ and $40\times 80$ are plotted on top of each-other. The data in (f) is only from the smaller system size, and the result is more tentative due to possible history dependency.

The lower the capillary number, the larger the role of capillary forces. Capillary forces may be sufficiently strong to hold bubbles of one fluid fixed in one place. However, for large capillary numbers all fluid is mobilised. After a while all relics of the initial fluid distribution is gone. We will make a subdivision of the parameter space into history independent and history dependent. Figures \ref{fig:capildep}(a)-(e) is in the history independent regime. By visual inspection we have observed that for capillary number $C_{\rm a}=1.0\times 10^{-4}$, pieces of the initial configuration can be seen all the time during the simulations. This also happens whenever the final situation is one where one fluid percolates and flows, while the other is immobile. If this were the final situation whatever initial configuration, it would not indicate history dependence. However, in Figure \ref{fig:capildep}(f) we can see very different results from almost the same saturation (at 40\% and at 60\%). It is in the history dependent regime. Here we also needed to simulate for a longer period of time to reach the steady-state. Further, general analysis of history dependent systems should be done with more care than we have done here, taking into account at least a number of possible initial configurations. Therefore, we classify this plot as tentative.

In the plots in Figure \ref{fig:capildep} a diagonal line is added. This is the line where the nonwetting fractional flow is equal to the nonwetting saturation. A miscible fluid mixture would follow this line. It is interesting to use this line as a reference and note how the data points lie above or below this line. Looking at the five plots \ref{fig:capildep}(a)-(e) in the history independent regime, we can say that (a) and (b) share the property that almost all data points are above the diagonal. In (d) and (e) the curves are S-shaped, while (c) makes an intermediate situation.

The plots (a) and (b) can be said to be in a viscous force-dominated regime. We can very well classify the history dependent regime as a capillary force-dominated regime. The regime in-between, plots (d) and (e), constitutes a viscous and capillary interactive regime.

In the viscous force-dominated regime the fractional flow curves are close to the diagonal. In the case of no capillary forces, no wetting difference, the fluids would have been interchangeable. This symmetry would have implied the fractional flow curve to be the straight line. However, there are capillary forces which statistically will make the wetting fluid occupy narrower tubes more frequently than is the case for the nonwetting fluid. Since the flow is faster in wider tubes, the fractional flow of nonwetting fluid is larger than the nonwetting saturation in this regime.

The interaction regime is more complex. The main aspects here are the distribution of the two fluids within pore space and the velocity field. For large nonwetting saturations the nonwetting fractional flow is larger than the saturation, and for small nonwetting saturation the fractional flow is smaller. Whenever one can find pathways through the network that contain few interfaces, they will in general carry more flux. There is a fundamental asymmetry in that the wetting fluid prefers being within tubes, while the nonwetting fluid prefers being around nodes. This implies that the fluid distribution is asymmetric. Our simulations show that larger compact wetting regions occur. Nonwetting structures are more branched, treelike, surrounding the wetting areas, generally speaking. A large wetting saturation implies that the pathway of least resistance passes through the compact wetting regions. Likewise for large nonwetting saturations the pathway is along the nonwetting branches. Briefly this explains qualitatively the S-shape of the curves in this regime.

\begin{figure}
\centering
\begin{tabular}{cc}
\includegraphics[height=5.0cm]{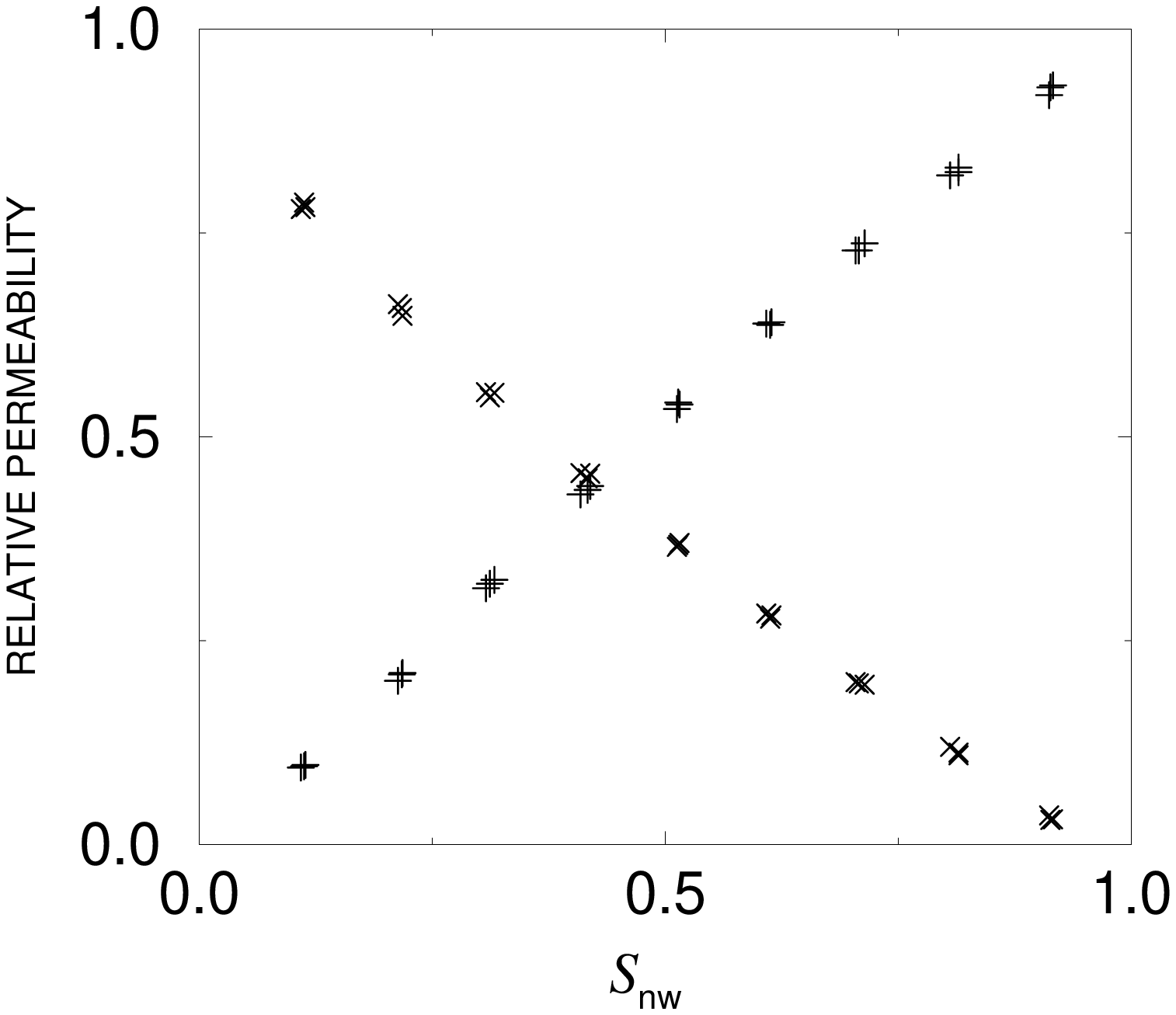} & 
\includegraphics[height=5.0cm]{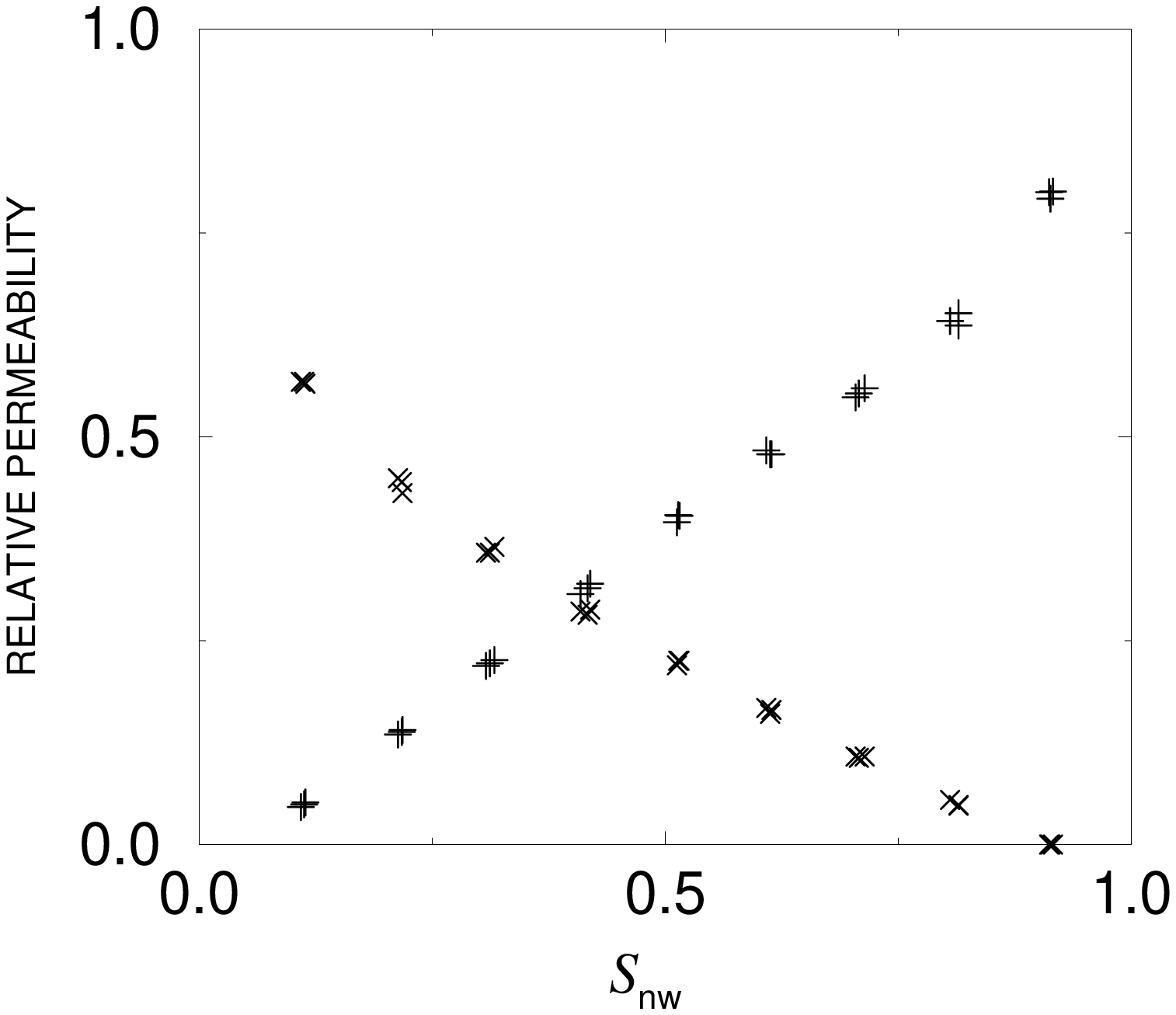} \\
(a) $C_{\rm a} = 3.2\times 10^{-2}$ &
(b) $C_{\rm a} = 1.0\times 10^{-2}$ \\  &  \\
\includegraphics[height=5.0cm]{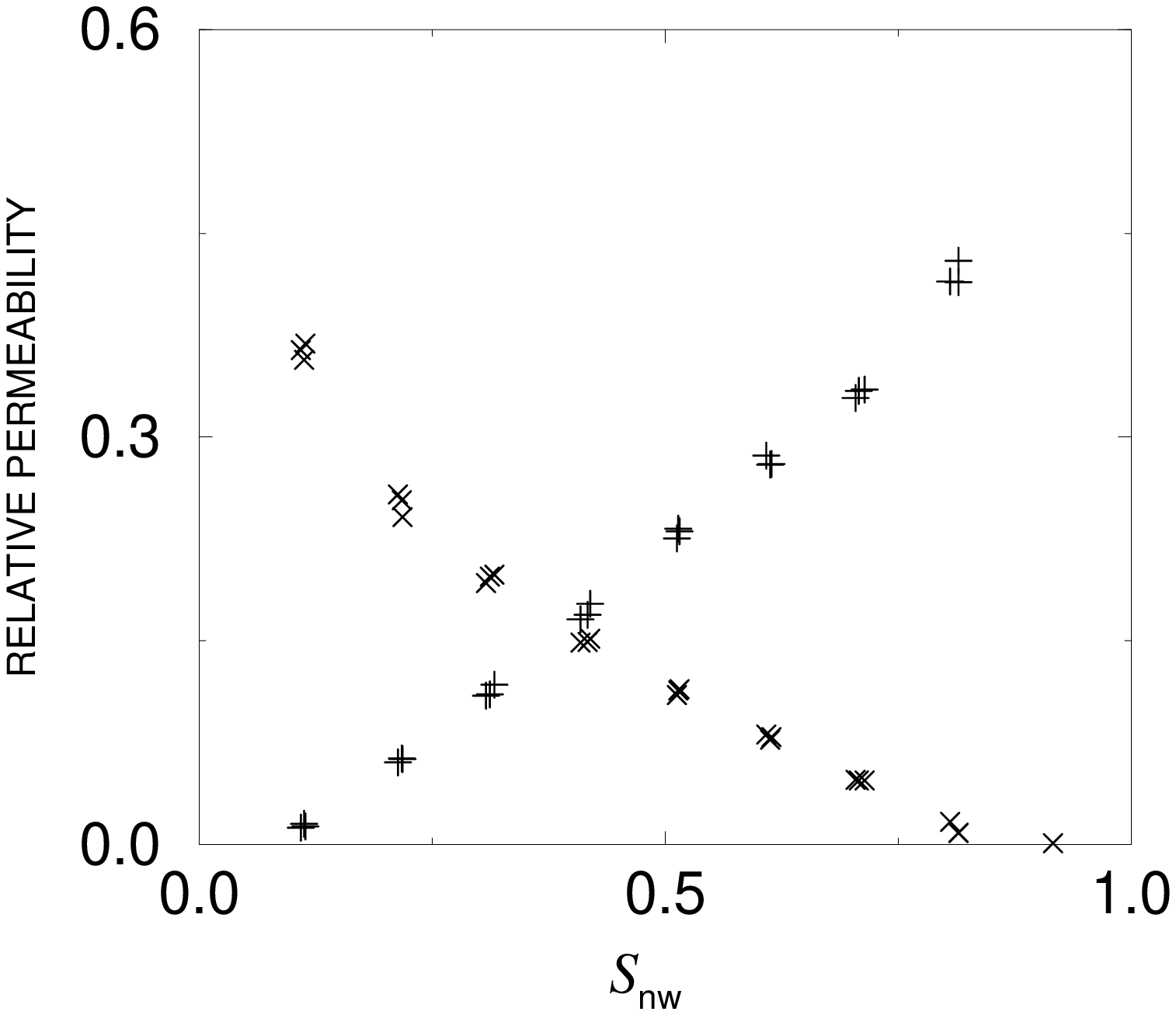} & 
\includegraphics[height=5.0cm]{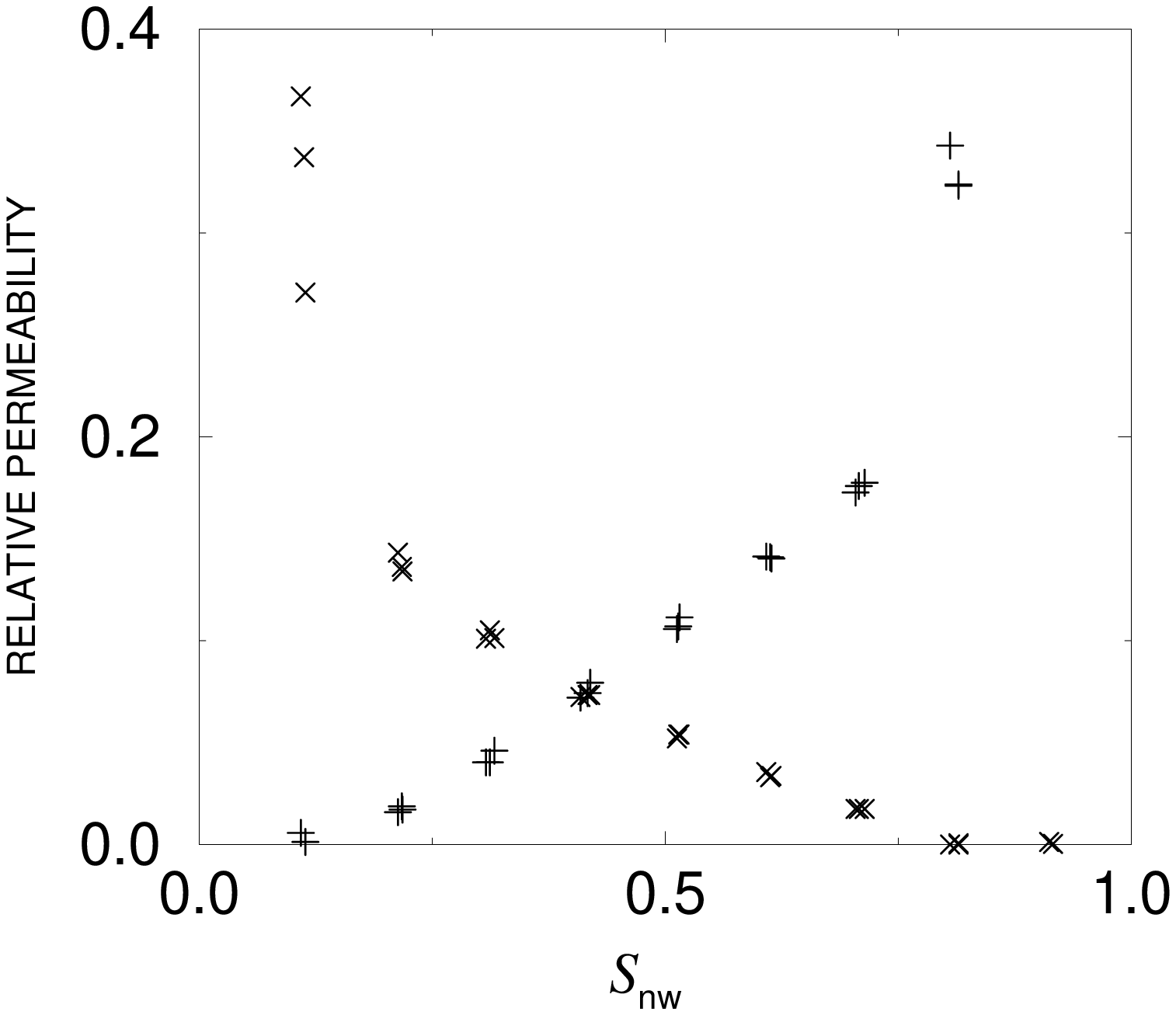} \\
(c) $C_{\rm a} = 3.2\times 10^{-3}$ &
(d) $C_{\rm a} = 1.0\times 10^{-3}$ \\  &  \\
\includegraphics[height=5.0cm]{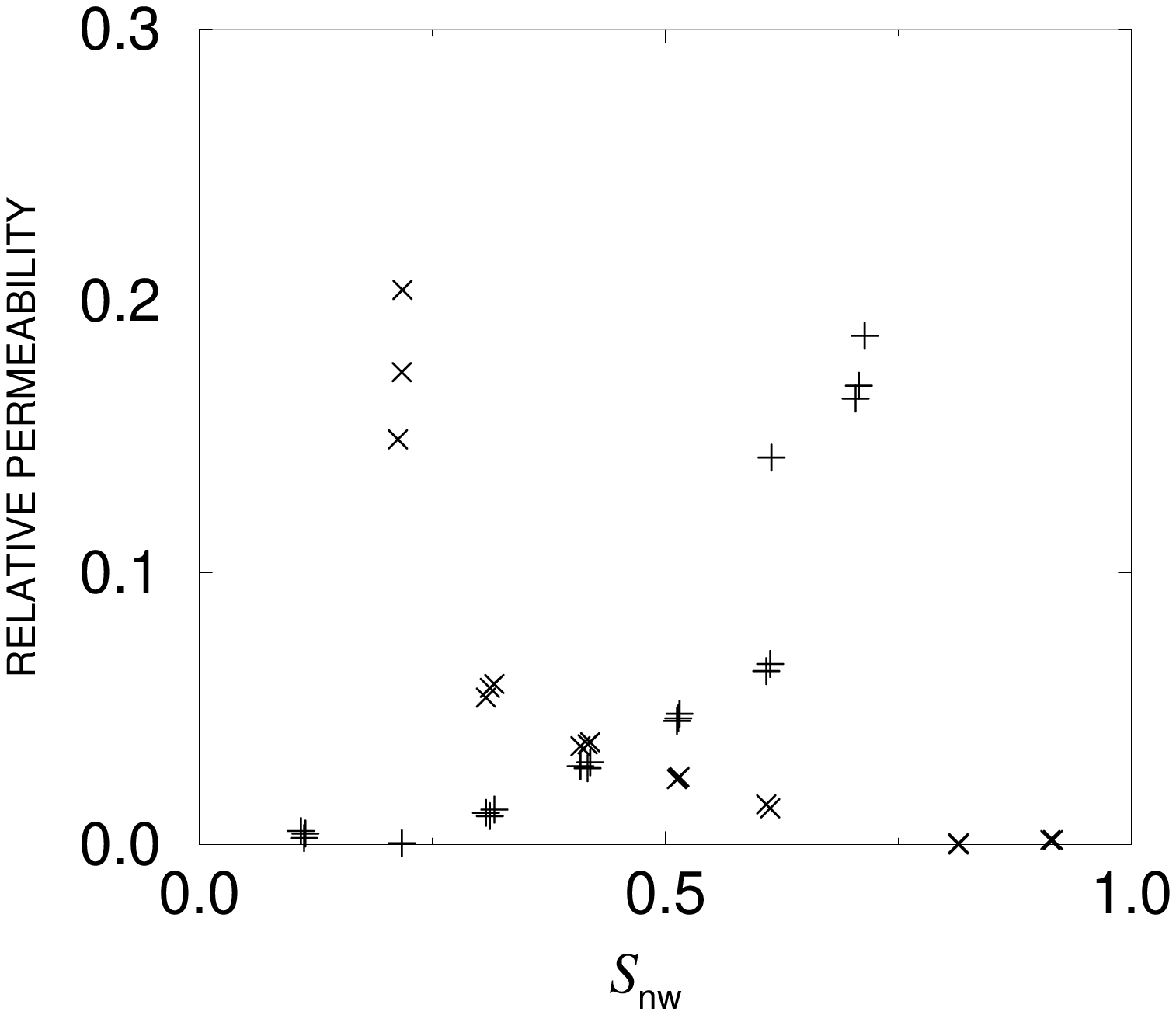} & 
\includegraphics[height=5.0cm]{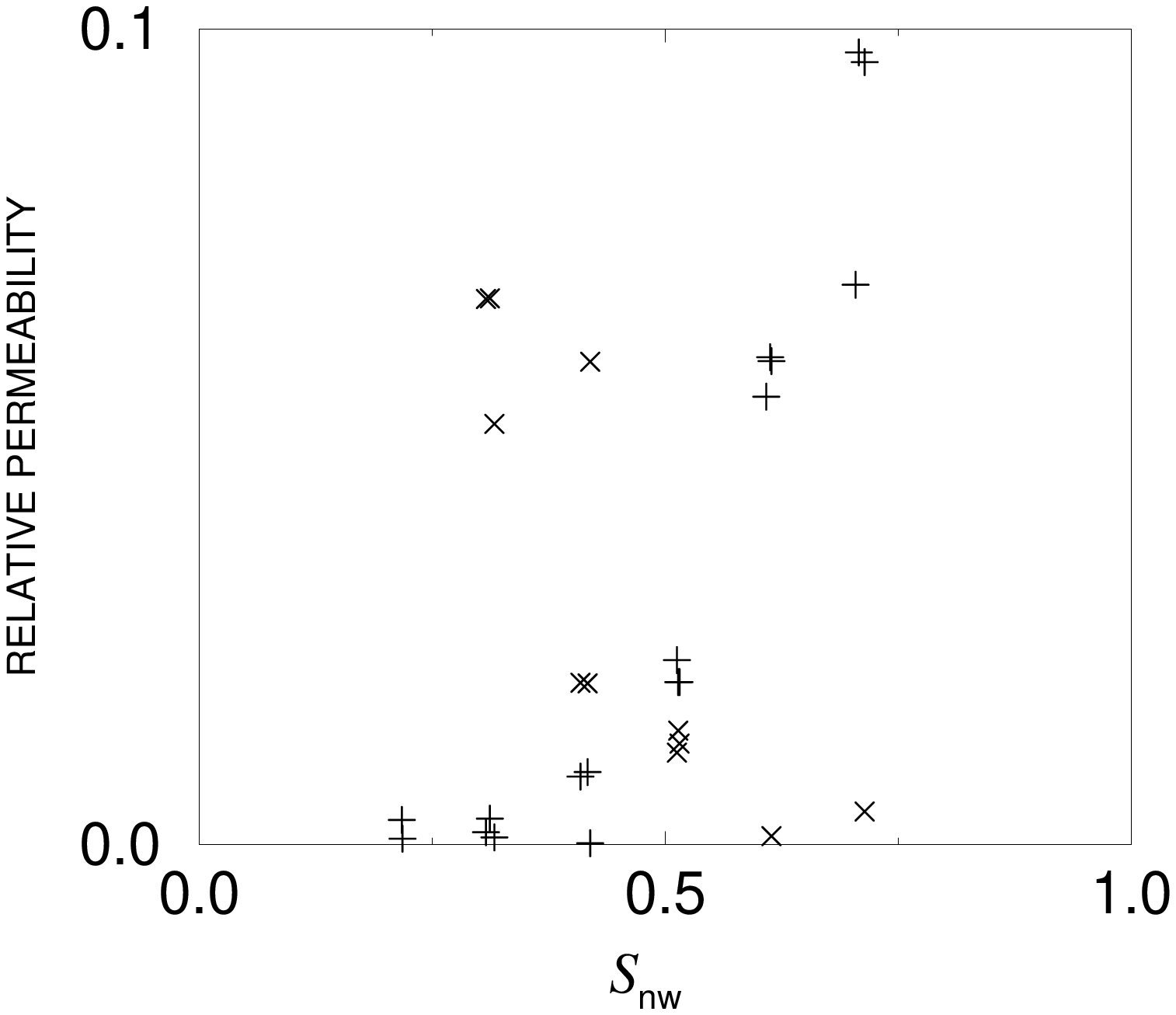} \\
(e) $C_{\rm a} = 3.2\times 10^{-4}$ &
(f) $C_{\rm a} = 1.0\times 10^{-4}$
\end{tabular}
\caption{\textsl{Nonwetting ($+$) and wetting ($\times$) relative permeabilities versus nonwetting saturation. The curves correspond to the fractional flow curves in Figure \ref{fig:capildep}. Also here (f) is tentative.}}
\label{fig:relperm}
\end{figure}

\subsection{Relative Permeabilities}
The fractional flow which was discussed in the previous subsection is only one of the interesting properties which can be studied with this network simulator. In this section we present the relative permeabilities, see Figure \ref{fig:relperm}, which correspond to the fractional flow curves in Figure \nolinebreak \ref{fig:capildep}.

The relative permeability formalism can be written as follows \cite{D92,HRAFJH97},
\begin{eqnarray}
\frac{Q_{\rm nw}(S_{\rm nw})}{\Sigma} =
\frac{k_{\rm r,nw}(S_{\rm nw})k}{\mu_{\rm nw}} \times
\frac{P(S_{\rm nw})}{L}, \\
\frac{Q_{\rm w}(S_{\rm nw})}{\Sigma} =
\frac{k_{\rm r,w}(S_{\rm nw})k}{\mu_{\rm w}} \times
\frac{P(S_{\rm nw})}{L}.
\end{eqnarray}
Here, $k_{\rm r,nw}$ and $k_{\rm r,w}$ is the nonwetting and wetting relative permeability, respectively. The globally applied pressure drop is $P$. The length of the system in the flow direction is $L$, and $k\ $ is the specific permeability of the network with respect to a single phase.

As it can be seen from Figure \ref{fig:relperm}, the sum of the two relative permeabilities are not unity;
\begin{equation}
k_{\rm r,nw}+k_{\rm r,w} \neq 1 .
\end{equation}
This deviation from unity is a result of the presence of capillary forces within the system. In Figure \ref{fig:relperm} the deviation increases from (a) to (f). This means that the effect of capillary forces increases, which in turn is caused by lowering of the capillary number.

Further, in Figure \ref{fig:relperm} (d)-(f) the nonwetting relative permeability is equal to zero, $k_{\rm r,nw}=0$, for nonwetting saturation below a certain threshold. Likewise, the wetting relative permeability is zero, $k_{\rm r,w}=0$, for nonwetting saturation above a different threshold.

\section{Concluding Remarks}
\label{sec:discussion}

The methodology is the main point of this article. We have presented a two-dimensional network simulator for two-phase flow. The model is based on a model which successfully has simulated drainage invasion \cite{AMHB98}. The major innovation is the introduction of {\it biperiodic boundary conditions}. This mimics conditions deep inside the reservoir. Concepts such as drainage and imbibition are fundamental in two-phase experiments. However, here they loose their meaning because it becomes impossible to tell which of the fluids is displacing the other and which is being displaced. The system is closed in the sense that the volume of the fluids is conserved. It is studied under a constant flux in one direction. This is done by explicit Euler time integration, and in each time step the appropriate global pressure, that gives the desired flux, is used.

By this method we have investigated the fractional flow properties of the system as a function of various parameters. These include the viscosities of the fluids, the total flux rate and the network size. The viscosity ratio and the capillary number are the important and effective parameters. We have also provided relative permeability curves which correspond to the fractional flow curves.

We have studied the case where the two fluids have equal viscosity, varying the capillary number. In particular, we found that for a fixed capillary number, the results were independent of the system size. The fact that fractional flow curves can be obtained by this method, simulating on reasonably small system sizes, makes it possible to obtain useful information with present computer power. Many questions need further investigation, the effect of changing the viscosity ratio being the most prominent.

The dependence on topology and tube and pore geometry is also a question that one must consider. The simulations are done in 2D which is qualitatively different from 3D. The big difference is that in 3D both the wetting and nonwetting phases can percolate the system. It is possible to have simultaneous flow of both fluids in separate parts of the system. This is not the case in 2D where flow of both fluids implies mixed pathways. For practical use of these kinds of simulations one needs to do 3D simulations. This is possible with the model as it is presented in this paper.

In our simulations in 2D we have used a regular topology, square network. The disorder is in the tube and pore geometry. The latter is characterised by the choice of letting $20\%$ of the tubes count in the effective pore volume. Also the distortions of the nodes, giving different tube lengths and the tube radius distribution, characterise this geometry. Our choices are all reasonable but not unique. Numerical tests have shown that the qualitative results are not sensitive to these details. However, we found it neccessary to fix these variables in order to understand the effect of other variables. If quantitative results for a specific medium are desired, one should probably reconsider these choices. In particular, the tube radius distribution is interesting to vary. Our preliminary results show that widening this distribution has a similar effect as lowering the capillary number. This effect is also larger for lower nonwetting saturation.

Based on the results presented here, we divided the parameter space into a division between history dependent and history independent regimes. The method works very well in the history independent regime. Whenever history dependence is important, care should be taken not to ignore possible impact of history on final results. Further, a division of the history independent regime was made into a viscous force-dominated regime and a viscous and capillary interactive regime. The latter being the more interesting having S-shaped fractional flow curves. These curves are comparable to the Buckley-Leverett type fractional flow curves. The simulator presented here constitutes an approach to obtaining this kind of data, that is different from the Buckley-Leverett approach.

\section*{Acknowledgements}
H.A.K., E.A. and A.H. thank the Niels Bohr Institute and Nordita for friendly hospitality and support. H.A.K. acknowledges support from VISTA, a collaboration between Statoil and The Norwegian Acad. of Science and Letters. E.A. acknowledges partial support from NFR through a SUP grant. We also like to thank Knut J{\o}rgen M{\aa}l{\o}y for valuable comments.

\bibliographystyle{klunamed}
\bibliography{referanser}

\begin{thebibliography}{}

\bibitem[\protect\citeauthoryear{Aker et~al.}{1998a}]{AMH98}
Aker, E., K.~J. M{\aa}l{\o}y, and A. Hansen: 1998a, `Simulating temporal
  evolution of pressure in two-phase flow in porous media'.
\newblock {\em Phys. Rev. E} {\bf 58}, 2217--2226.

\bibitem[\protect\citeauthoryear{Aker et~al.}{1998b}]{AMHB98}
Aker, E., K.~J. M{\aa}l{\o}y, A. Hansen, and G.~G. Batrouni: 1998b, `A
  Two-Dimensional Network Simulator for Two-Phase Flow in Porous Media'.
\newblock {\em Transport in Porous Media} {\bf 32}, 163--186.

\bibitem[\protect\citeauthoryear{Avraam and Payatakes}{1995a}]{AP95a}
Avraam, D.~G. and A.~C. Payatakes: 1995a, `Flow regimes and relative
  permeabilities during steady-state two-phase flow in porous media'.
\newblock {\em J. Fluid Mech.} {\bf 293}, 207--236.

\bibitem[\protect\citeauthoryear{Avraam and Payatakes}{1995b}]{AP95b}
Avraam, D.~G. and A.~C. Payatakes: 1995b, `Generalized Relative Permeability
  Coefficients during Steady-State Two-Phase Flow in Porous Media, and
  Correlation with the Flow Mechanisms'.
\newblock {\em Transport in Porous Media} {\bf 20}, 135--168.

\bibitem[\protect\citeauthoryear{Avraam and Payatakes}{1999}]{AP99}
Avraam, D.~G. and A.~C. Payatakes: 1999, `Flow Mechanisms, Relative
  Permeabilities, and Coupling Effects in Steady-State Two-Phase Flow through
  Porous Media. The Case of Strong Wettability'.
\newblock {\em Ind. Eng. Chem. Res.} {\bf 38}, 778--786.

\bibitem[\protect\citeauthoryear{Batrouni and Hansen}{1998}]{BH98}
Batrouni, G.~G. and A. Hansen: 1998, `Fracture in Three-Dimensional Fuse
  Networks'.
\newblock {\em Phys. Rev. Lett.} {\bf 80}, 325--328.

\bibitem[\protect\citeauthoryear{Binning and Celia}{1998}]{BC98}
Binning, P. and M.~A. Celia: 1998, `Practical implementation of the fractional
  flow approach to multi-phase flow simulation'.
\newblock {\em Advances in Water Resources} {\bf 22}, 461--478.
\newblock and references herein.

\bibitem[\protect\citeauthoryear{Blunt and King}{1990}]{BK90}
Blunt, M. and P. King: 1990, `Macroscopic parameters from simulations of pore
  scale flow'.
\newblock {\em Phys. Rev. A} {\bf 42}, 4780--4787.

\bibitem[\protect\citeauthoryear{Buckley and Leverett}{1942}]{BL42}
Buckley, S.~E. and M.~C. Leverett: 1942.
\newblock {\em Trans. Am. Inst. Min. Eng.} {\bf 146}, 107.

\bibitem[\protect\citeauthoryear{Chen and Wilkinson}{1985}]{CW85}
Chen, J.-D. and D. Wilkinson: 1985, `Pore-scale viscous fingering in porous
  media'.
\newblock {\em Phys. Rev. Lett.} {\bf 55}, 1892--1895.

\bibitem[\protect\citeauthoryear{Cieplak and Robbins}{1990}]{CR90}
Cieplak, M. and M.~O. Robbins: 1990, `Influence of contact angle on quasistatic
  fluid invasion of porous media'.
\newblock {\em Phys. Rev. B} {\bf 41}, 11508--11521.

\bibitem[\protect\citeauthoryear{Dullien}{1992}]{D92}
Dullien, F. A.~L.: 1992, {\em Porous Media: Fluid Transport and Pore
  Structure}.
\newblock Academic Press, San Diego.

\bibitem[\protect\citeauthoryear{Haines}{1930}]{H30}
Haines, W.~B.: 1930, `Studies in the physical properties of soil'.
\newblock {\em J. Agr. Sci.} {\bf 20}, 97--116.

\bibitem[\protect\citeauthoryear{Hansen et~al.}{1997}]{HRAFJH97}
Hansen, A., S. Roux, A. Aharony, J. Feder, T. J{\o}ssang, and H.~H. Hardy:
  1997, `Real-Space Renormalization Estimates for Two-Phase Flow in Porous
  Media'.
\newblock {\em Transport in Porous Media} {\bf 29}, 247--279.

\bibitem[\protect\citeauthoryear{Koplik and Lasseter}{1985}]{KL85}
Koplik, J. and T.~J. Lasseter: 1985, `Two-phase flow in random network models
  of porous media'.
\newblock {\em SPE J} {\bf 2}, 89--100.

\bibitem[\protect\citeauthoryear{Lenormand et~al.}{1988}]{LTZ88}
Lenormand, R., E. Touboul, and C. Zarcone: 1988, `Numerical models and
  experiments on immiscible displacements in porous media'.
\newblock {\em J. Fluid Mech.} {\bf 189}, 165--187.

\bibitem[\protect\citeauthoryear{Lenormand and Zarcone}{1983}]{LZ83}
Lenormand, R. and C. Zarcone: 1983, `Mechanism of the displacement of one fluid
  by another in a network of capillary ducts'.
\newblock {\em J. Fluid Mech.} {\bf 135}, 337--353.

\bibitem[\protect\citeauthoryear{M{\aa}l{\o}y et~al.}{1985}]{MFJ85}
M{\aa}l{\o}y, K.~J., J. Feder, and T. J{\o}ssang: 1985, `Viscous fingering
  fractals in porous media'.
\newblock {\em Phys. Rev. Lett.} {\bf 55}, 2688--2691.

\bibitem[\protect\citeauthoryear{M{\aa}l{\o}y et~al.}{1992}]{MFFJ92}
M{\aa}l{\o}y, K.~J., L. Furuberg, J. Feder, and T. J{\o}ssang: 1992, `Dynamics
  of slow drainage in porous media'.
\newblock {\em Phys. Rev. Lett.} {\bf 68}, 2161--2164.

\bibitem[\protect\citeauthoryear{Paterson}{1984}]{P84}
Paterson, L.: 1984, `Diffusion-limited aggregation and two-fluid displacements
  in porous media'.
\newblock {\em Phys. Rev. Lett.} {\bf 52}, 1621--1624.

\bibitem[\protect\citeauthoryear{Rothman}{1990}]{R90}
Rothman, D.~H.: 1990, `Macroscopic laws for immiscible 2-phase flow in
  porous-media - results from numerical experiments'.
\newblock {\em J Geophys. Res.} {\bf 95}, 8663--8674.

\bibitem[\protect\citeauthoryear{Roux}{unpublished}]{roux}
Roux, S.: unpublished.

\bibitem[\protect\citeauthoryear{Washburn}{1921}]{W21}
Washburn, E.~W.: 1921, `The dynamics of capillary flow'.
\newblock {\em Phys. Rev.} {\bf 17}, 273--283.

\bibitem[\protect\citeauthoryear{Wilkinson and Willemsen}{1983}]{WW83}
Wilkinson, D. and J.~F. Willemsen: 1983, `Invasion percolation: A new form of
  percolation theory'.
\newblock {\em J. Phys. A} {\bf 16}, 3365--3376.

\bibitem[\protect\citeauthoryear{Witten and Sander}{1981}]{WS81}
Witten, T.~A. and L.~M. Sander: 1981, `Diffusion-limited aggregation, a kinetic
  critical phenomenon'.
\newblock {\em Phys. Rev. Lett.} {\bf 47}, 1400--1403.

\end{thebibliography}

\end{article}
\end{document}